\documentclass{article}

\date{\today}

\usepackage{graphicx}
\usepackage[symbol]{footmisc}
\usepackage{amsmath, amssymb, amsthm, amsfonts}
\usepackage{color}
\usepackage{lineno}
\usepackage{mathtools}
\usepackage{subcaption}
\usepackage{enumitem}
\usepackage{comment}
\usepackage{tikz}
\usepackage[algoruled,vlined]{algorithm2e}
\usetikzlibrary{arrows, positioning, cd}
\usepackage{hyperref}
\hypersetup{
    colorlinks = true,
    allcolors = blue
}

\newcounter{problem}

\newtheorem{theorem}{Theorem}
\newtheorem{remark}{Remark}

\newtheorem{corollary}[theorem]{Corollary}

\newtheorem{lemma}[theorem]{Lemma}

\newtheorem{prob}[problem]{Problem}

\newtheoremstyle{case}{}{}{}{}{}{:}{ }{}
\theoremstyle{case}

\def\OptSol{{\sc OptSol}}
\def\AppSol{{\sc AppSol}}
\def\P{\mathcal{P}}

\def\CH{\mbox{{\it CH}}}
\def\Int{\mbox{\em Int}}
\def\Ext{\mbox{\em Ext}}
\def\modu{\mbox{\em mod }}
\def\modr{\mbox{mod }}
\def\id#1{\mbox{\emph{#1}}}

\catcode`\@=11
\long\def\@makecaption#1#2{
 \vskip 10pt
 \setbox\@tempboxa\hbox{\bf #1: \sf #2}
 \ifdim \wd\@tempboxa >\hsize \bf #1: \sf #2\par \else \hbox
to\hsize{\hfil\box\@tempboxa\hfil}
 \fi}
\catcode`\@=12

\providecommand{\keywords}[1]{\textbf{\textit{Keywords---}} #1}

\begin{document}

\title{
Optimal Placement of Base Stations in Border Surveillance using Limited Capacity Drones}

\author{S. Bereg\thanks{Department of Computer Science, University of Texas at Dallas, Richardson, TX 75080, USA. {\tt besp@utdallas.edu, Mohammadreza.Haghpanah@utdallas.edu}.  Supported in part by NSF award CCF-1718994.} 
\and J.M. D\'iaz-B\'a\~nez 
\thanks{Departamento de Matem\'atica Aplicada II, Universidad de Sevilla, Spain, {\tt dbanez@us.es, frodriguex@us.es}. Partially supported by European Union's Horizon 2020 research and innovation programme under the Marie Sk\l{}odowska-Curie grant agreement \#734922 and Ministerio de Ciencia e Innovación CIN/AEI/10.13039/501100011033/(PID2020-114154RB-I00).}
\and M. Haghpanah$^*$ 
\and P. Horn\thanks{Department of Mathematics, University of Denver. {\tt paul.horn@du.edu, alex.stevens@du.edu}.  Supported in part by Simons Collaboration Grant \#525039}  \and M.A. Lopez %
\thanks{Department of Computer Science, University of Denver, {\tt mario.lopez@du.edu}. Supported in part by a University of Denver John Evans Research Award.}
\and N. Marín\thanks{Corresponding author. Posgrado en Ciencia e Ingeniería de la Computación, Universidad Nacional Aut\'onoma de M\'{e}xico. {\tt nestaly@ciencias.unam.mx}. ORCID: 0000-0002-0222-3254. 
Supported by SEP-CONACyT of Mexico.}%
\and A. Ramírez-Vigueras\thanks{Instituto de Matemáticas, UNAM. {\tt adriana.rv@im.unam.mx}. Partially supported by PAPIIT IN105221, Programa de Apoyo a la Investigación e Innovación Tecnologíca UNAM.}%
\and F. Rodr{\'\i}guez $^\dag$ %\thanks{Departamento de Matem\'atica Aplicada II, Universidad de Sevilla, Spain, {\tt frodriguex@us.es}.}
\and O. Solé-Pi \thanks{Departamento de Matemáticas, Facultad de Ciencias, UNAM. \mbox{{\tt oriolandreu@ciencias.unam.mx}}.}% 
\and A. Stevens$^\ddagger$
\and J. Urrutia \thanks{Instituto de Matem\'{a}ticas, Universidad Nacional Aut\'onoma de M\'{e}xico, {\tt urrutia@matem.unam.mx}. ORCID: 0000-0002-4158-5979. Supported in part by PAPIIT IN102117 Programa de Apoyo a la Investigaci\'on e Innovaci\'on Tecnol\'ogica, UNAM.}
}

\maketitle
\begin{abstract}
Imagine an island modeled as a simple polygon $\P$ with $n$ vertices whose coastline we wish to monitor. 
We consider the problem of building the minimum number of refueling stations along the boundary of $\P$ in such a way that a drone can follow a polygonal route enclosing the island without running out of fuel. 
A drone can fly a maximum distance $d$ between consecutive stations and is restricted to move either along the boundary of $\P$ or its exterior (i.e., over water).
We present an algorithm that, given $\mathcal P$, finds the locations for a set of refueling stations whose cardinality is at most the optimal plus one.
The time complexity of this algorithm is $O(n^2 + \frac{L}{d} n)$, where $L$ is the length of $\mathcal P$.
We also present an algorithm that returns an additive $\epsilon$-approximation for the problem of minimizing the fuel capacity required for the drones when we are allowed to place $k$ base stations around the boundary of the island; this algorithm also finds the locations of these refueling stations. 
Finally, we propose a practical discretization heuristic which, under certain conditions, can be used to certify optimality of the results. 
\end{abstract}


\section{Introduction}

 The rapid development and use of \emph{Unmanned Aerial Vehicles} (UAVs), commonly called drones, in many activities of our daily life has created a need for the development of new algorithms to optimize their use. 
 A factor that is common to most types of drones is their relatively short flying range due mostly to their restricted energy capacity \cite{long2018energy}.

Border patrolling is one typical application of air surveillance systems where the deployment of drones has become a natural choice for providing monitoring, surveillance, and search and rescue services for the protection of human lives or natural resources~\cite{alzahrani2020uav, karaca2018potential, li2018unmanned,merwaday2016improved, ribeiro2021unmanned}.
In this context, a team of UAVs can be deployed along the boundaries of a region to collect useful information, such as images or videos, and send it to the nearest control center. The energy limitation of small UAVs prevents them from remaining in flight for long periods of time. 
Thus, recharging stations, platforms where the drone can
autonomously land to recharge its battery before continuing its mission, have been recently introduced. However, the cost of those platforms remains a significant obstacle and, consequently, it is important to reduce their number.
 
\begin{figure}[h!]
    \begin{center}
        \begin{tikzpicture}[scale=0.3]
            \draw [color=darkgray] (15.,10.5)-- (12.,9.5)-- (12.,3.)-- (15.5,1.5)-- (11.,1.5)--(10.,4.5)-- (3.5,4.5)-- (2.,1.)-- (2.,5.5)-- (5.,6.5)--(5.,13.)-- (1.5,14.5)-- (6.,14.5)-- (7.,11.5)--(13.5,11.5)-- (15.,15.)--(15.,10.5);
            
            \draw [line width=1.2pt,dash pattern=on 4pt off 4pt,color=black] (6.,14.5)-- (10.25,11.5)-- (15.,15.)-- (15.,12.75)-- (15.,10.5)-- (12.,6.25)-- (15.5,1.5)-- (13.25,1.5)-- (11.,1.5)-- (6.75,4.5)-- (2.,1.)-- (2.,3.25)-- (2.,5.5)-- (5.,9.75)-- (1.5,14.5)--(3.75,14.5)-- (6.,14.5);
         
            \draw [fill=black] (5.,9.75) circle (7pt);
            \draw [fill=black] (10.25,11.5) circle (7pt);
            \draw [fill=black] (12.,6.25) circle (7pt);
            \draw [fill=black] (6.75,4.5) circle (7pt);
            \draw [fill=black] (15.,12.75) circle (7pt);
            \draw [fill=black] (13.25,1.5) circle (7pt);
            \draw [fill=black] (2.,3.25) circle (7pt);
            \draw [fill=black] (3.75,14.5) circle (7pt);
            
            %\draw (current bounding box.south east) rectangle (current bounding box.north west);
        \end{tikzpicture}
        \caption{Example. A polygon $\mathcal P$ and $d$-hull $\mathcal C$ (black dashed line) that uses an optimal set of base stations (solid circles) for the MinStation problem.}
        \label{fig:island1}
    \end{center}
\end{figure}
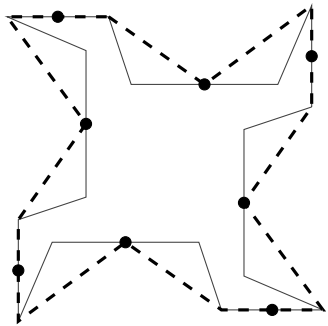

Inspired by this type of applications, we study in this paper the following geometric optimization problems: 

\emph{The MinStation Problem}:
 Suppose that we want to guard the border of an island $\mathcal I$ whose boundary is modeled by a simple polygon $\mathcal P$ using a set of drones that can fly a distance $d$ before they need refueling.
 Our drones can fly over the boundary of $\mathcal I$ or over small sectors of the sea surrounding it, but not over the interior of $\mathcal I$. 
 Our objective is to place a set $S=\{s_0, \ldots , s_{k-1}\}$ of $k$ refueling base stations with \emph{minimum cardinality $k$} and located on the boundary of $\mathcal I$, such that when a drone visits all the refueling stations it travels a closed curve that encloses $\mathcal P$; the flying distance between $s_i$ and $s_{i+1}$ is at most $d$, with addition taken $\modr k$. 
 See Figure~\ref{fig:island1} for an example. 
 We will refer to $S$ as an \emph{optimal solution}. 
 A set $S'=\{s'_0, \ldots , s'_k\}$ with $k+1$ refueling stations will be called a \emph{quasi-optimal} solution.
 
\emph{The MinDistance Problem}: Suppose that we have a budget that allows us to build $k$ refueling stations. 
Find the smallest $d$ such that we can build $k$ refueling stations that allow a drone with flight capacity $d$ to guard the border of the given island.  
 
Solutions to these problems can be implemented using fleets of small drones with limited capacity, resulting in cheaper systems that use less resources and increase the frequency with which the drones patrol the border.
We consider that a drone can replace the used battery with a fresh one, rather than charging the battery on site. 
The used batteries can be recharged when a charging station is not servicing drones. 
Thus, a drone will not occupy a charging station for a long time and there is at most one drone at each station at the same time.
Our main contributions are as follows:

\medskip
\begin{itemize}

\item We give an algorithm, \OptSol, with complexity $O(n^2 + \frac{L}{d} n)$, such that if $s_0$ is a fixed point on the convex hull of $\mathcal P$, $\CH(P)$, finds an optimal solution $S$ to the MinStation Problem under the restriction that $s_0 \in S$; here, $L$ is the perimeter of $\mathcal P$.  
This yields either an optimal or quasi-optimal solution to the unconstrained MinStation Problem (without requiring $s_0\in S$).
The problem of finding an optimal unconstrained solution is equivalent to that of finding the location of a single station in an optimal solution. 
We leave as an open problem that of designing a polynomial time algorithm for the general unconstrained case.

\item For the \emph{MinDistance Problem}, we show how to approximate an optimal solution up to an additive constant.
The approach is based on an algorithm, \emph{AppSol}, which solves a discretized version of the MinStation Problem.

\item We implemented \emph{AppSol} and ran experiments on a polygon corresponding to an island (using GIS data); in many cases, the algorithm returns a certifiably optimal solution on this real data.

\end{itemize}

\subsection{Related work}
Drones have become the natural choice for the deployment of air surveillance systems~\cite{alzahrani2020uav,li2018unmanned,manyam2017multi}. 
We mention three areas where problems close to ours can be found: facility location, wireless networks, and computational geometry.

In facility location problems we are interested in finding the best places to locate a set of resources (e.g. airports, pharmacies, gas stations, markets, etc.) to better serve a community, as well as creating optimal routes to visit them. 
The facilities can be isolated points in space~\cite{laporte2019introduction} or 2-dimensional structures such as straight lines, line-segments, polygonal curves, or circles ~\cite{diaz2004continuous}.
The location-routing problem is a research area within locational analysis, with the distinguishing property of paying special attention to vehicle routing aspects~\cite{nagy2007location}. 
Recently, applications in the aerial robotics community, such as finding the best places to locate drone base stations and creating flying routes for the drones, have arisen in areas such as border patrolling \cite{saricciccek2015unmanned}.
Cities on the borders of countries are modeled as demand points, and airports are considered as base stations or hubs. 
In~\cite{yakici2016solving} the authors studied a base location and path planning problem in maritime target reconnaissance problems.  
Their problem is formulated as an integer linear program where the total score obtained from visiting points of interest by flight routes of drones is maximized; a novel ant colony optimization metaheuristic approach is proposed.
In a more recent paper \cite{liu2019optimization}, both capacity constraints on base stations and endurance limitations on drones are taken into account and two heuristic algorithms are designed to solve the problem.
A similar problem is considered in \cite{cicek2019location}. 
They investigate the 3D location problem of multiple drone base stations as well as the allocation of their dynamic capacities to the users. 
The service provided by the stations is also dynamic in terms of the data rate level provided to the users. 
In addition to border patrolling, the problem of deploying a number of charging stations to cover a demand region has been considered in several areas. 
For example, \cite{hong2018range} proposes a coverage model to figure out the optimal positions of a given number of charging stations from a discretized candidate set, with the objective of maximizing the coverage of customers. 
More recently, a connectivity requirement on the stations is required in \cite{huang2020method} in order to guarantee the delivery to customers located far from a depot. 
They argue that any two neighbor charging stations should be within a certain range such that a drone with a fully charged battery can reach one from the other. 
In other words, it is required that the deployed charging stations should be connected to the depot. %\red{What does neighbor mean here and what does it mean to be connected to a depot?}
Then, a drone that departs from the depot can arrive at any charging station via a subset of other charging stations, and it can then service customers near this charging station.

Another field of research close to our work can be found in wireless sensor networks. 
In~\cite{kumar2007barrier}, the authors study the $k$-barrier problem: how to deploy a set of sensors in a belt region surrounding a castle in such a way that any intruder is detected by at least $k$ sensors.
In \cite{bhattacharya2009optimal} the following problem is studied: Given that an intruder has been detected by a set of sensors, how can they be moved in an optimal way to prevent further intrusions? 
An interesting survey of problems similar to ours can be found in \cite{ahmed2005holes}, where they study the problem of protecting several types of holes that can occur in a wireless sensor network, where a hole is a region not covered by the sensing disks of a set of sensors. In the same paper, other problems related to ours are considered, including routing in static and mobile sensor networks.
See also~\cite{bose2001routing,ganeriwal2004self,wang2006movement}.

Finally, in computational geometry, there is a whole area of research devoted to problems, collectively known as Art Gallery problems, which are related to ours. 
The oldest problem in this area requires finding a minimum set of points $S$ within an art gallery, usually modeled by a polygon $\mathcal P$, such that every point in the polygon is visible from at least one point in $S$.  
Many variations of this first problem have since been studied and the interested reader is referred to the book by O'Rourke~\cite{o1987art} or the surveys by Shermer~\cite{shermer1992recent} and Urrutia~\cite{urrutia2000art}.
Some variations of the art gallery problem, more closely related to our problem, consider the use of mobile guards. 
The \emph{Watchman Route problem}, introduced in~\cite{chin1986optimum}, is that of finding a path of minimum length that a guard can follow in order to guarantee that every point within $\mathcal P$ is visible from some point in the path. 
There are two main variations of this problem, one in which the starting point of the route is given~\cite{tan1999corrigendum}, as well as the unrestricted case~\cite{tan2001fast}.
Even more closely related to our problem is that of finding the minimum watchman route in the exterior of $\P$ \cite{ntafos1994external}.
Some variations of the watchman route problem require the path to visit a set of $k$ sites represented as polygons in $\P$. 
In the \emph{Safari Route problem}~\cite{tan2003finding} we are allowed to enter the sites, while in the \emph{Zoo Keeper Route problem}~\cite{wei1992zookeeper, bespamyatnikh2003nlogn} the guard has to visit their boundary but is not allowed to go inside (as when feeding an animal without entering its cage).
Both problems are NP-hard in the general case, but can be solved in polynomial time if the sites are adjacent to the boundary of $\P$.
The \emph{Aquarium Keeper's problem}, studied in~\cite{czyzowicz1991aquarium}, deals with the problem of computing the shortest closed path inside $\P$ which visits each of its edges at least once.

\section{Terminology and Problem Formulation}

In what follows a polygon $\mathcal P$ is represented by a sequence $\langle p_0, \ldots, p_{n-1}\rangle$ of its vertices given in clockwise order around its boundary. 
Thus, the edges of $\mathcal P$ are the line segments $\overline{p_ip_{i+1}}$, with addition taken $\modr n$. 
We assume that our polygons are always \emph{simple}, i.e. that no two non-consecutive edges intersect. 
We use $\Int(\mathcal P)$ and $\Ext(\mathcal P)$ to denote, respectively, the interior and exterior of the region enclosed by $\mathcal P$, and use $\mathcal P$ itself to refer to the boundary of this region (often referred to by $\partial{\mathcal P}$ in the literature). 
Accordingly, the length of $\mathcal P$ is the sum of the lengths of its edges. 
A (polygonal) path is a sequence of points $\langle q_0, \ldots, q_{k}\rangle$ together with the set of edges $\overline{q_{i} q_{i+1}}$, $i=0,\ldots,k-1$; the length of a path is the sum of the lengths of its edges.

Given two distinct points $a, b \in \mathcal P$ the interval $[a,b]$ is the set of points of $\mathcal P$ traversed while moving from $a$ to $b$ in the clockwise direction along the boundary of $\mathcal P$. 
The distance $\delta_\mathcal P(a,b)$ between $a$ and $b$  in $\mathcal P$ is the length of the interval $[a,b]$. 
Observe that since $a \neq b$, $[a,b] \neq [b,a]$, $[a,b] \cup [b,a] = \mathcal P$, and that $\delta_\mathcal P(a,b) + \delta_\mathcal P(b,a)$ is the length of $\mathcal P$. 

The following definitions of what we will call $d$-paths and $d$-hulls arise from the restriction that the flight range of a drone is a fixed number $d$. 

An open line segment contained in $\Ext(\mathcal P)$ joining two points $a,b \in \mathcal P$ will be called \emph{a bridge}; if its length is at most $d$ it is called a $d$-bridge of $\mathcal P$.
Note that a drone cannot fly along a bridge of $\mathcal P$ with length greater than $d$; the base stations are restricted to be on $\mathcal P$, thus if a drone chooses to fly over a bridge with length greater than $d$ it would run out of fuel and fall to the sea. 

A polygonal path joining two points $a,b \in \mathcal P$ is called a $d$-path if all of its edges are $d$-bridges of $\mathcal P$, or segments of edges of $\mathcal P$. 
We say that a polygon $\mathcal C$ is a $d$-hull of $\mathcal P$ if it encloses $\mathcal P$, and all of its edges are contained in edges of $\mathcal P$ or are $d$-bridges of $\mathcal P$. 
Observe that a polygon has many (in fact an infinite number) of $d$-hulls, indeed $\mathcal P$ is a $d$-hull of itself.

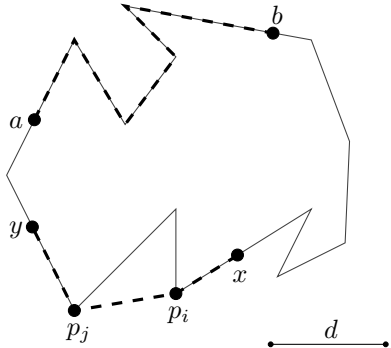
\begin{figure}[htb]
    \begin{center}
        \begin{tikzpicture}[scale=0.45]
            \draw [color=darkgray] (3.,2.)-- (1.,6.)-- (3.,10.)-- (4.5,7.5)-- (6.,9.5)-- (4.5,11.)-- (10.,10.)-- (11.13,7)-- (11.,4.)-- (9.,3.)-- (10.,5.)-- (6.,2.5)-- (6.,5.)-- (3.,2.);
            
            \draw [] (8.8,1)-- (12.2,1);
            \draw [fill=black] (8.8,1) circle (2pt);
            \draw [fill=black] (12.2,1) circle (2pt);
            \draw (10.1,2) node[anchor=north west] {$d$};
            
            \draw [line width=1.2pt,dash pattern=on 4pt off 4pt,color=black] (7.82,3.63)-- (6.,2.5)-- (3.,2.)-- (1.76,4.47);
            
            \draw [line width=1.2pt,dash pattern=on 4pt off 4pt,color=black] (1.82,7.64)-- (3.,10.)-- (4.5,7.5)-- (6.,9.5)-- (4.5,11.)-- (8.87,10.2);
            
            \draw (2.5,1.9) node[anchor=north west] {$p_j$};
            \draw (5.5,2.4) node[anchor=north west] {$p_i$};
            \draw (0.8,8) node[anchor=north west] {$a$};
            \draw (8.55,11.3) node[anchor=north west] {$b$};
            \draw (7.4,3.45) node[anchor=north west] {$x$};
            \draw (0.8,4.9) node[anchor=north west] {$y$};
            
            \draw [fill=black] (3.,2.) circle (5pt);
            \draw [fill=black] (6.,2.5) circle (5pt);
            \draw [fill=black] (1.76,4.47) circle (5pt);
            \draw [fill=black] (7.82,3.64) circle (5pt);
            \draw [fill=black] (1.82,7.64) circle (5pt);
            \draw [fill=black] (8.87,10.20) circle (5pt);
        
            %\draw (current bounding box.south east) rectangle (current bounding box.north west);
        \end{tikzpicture}
    \caption{The interval $[a, b] $ and a $d$-path $\pi_{x, y}$ joining $x$ and $y$ are shown in black dashed lines. The segment $\overline{p_i p_j}$ is a $d$-bridge contained in $\pi_{x, y}$.}
    \label{fig:definitions}
    \end{center}
\end{figure}

The \emph{drone distance} $\delta(a,b)$ from $a$ to $b$ is the length of a shortest clockwise $d$-path joining $a$ to $b$. As an example, Figure~\ref{fig:definitions} shows the shortest $d$-path from $x$ to $y$. 
For simplicity, we will refer to the drone distance as the distance from $a$ to $b$. Observe that $\delta(a,b)$ is in general different from $\delta(b,a)$. 
Further observe that the drone distance and the geodesic distance from $a$ to $b$ (understood as the length of the shortest clockwise path from $a$ to $b$ disjoint from $\Int(\mathcal P)$) coincide whenever one of them is at most $d$.
Finally, note that if a drone with flight range $d$ can fly between two points $a,b \in \mathcal P$ without recharging, then there is a $d$-path joining them of length at most $d$.

Our island guarding problem can now be restated as follows:

\begin{prob}[{MinStation}] Given a polygon $\mathcal P$ find a set of base stations $S=\{s_0, \ldots , s_{k-1}\}$ with minimum cardinality such that for every $0 \leq i \leq k-1$ there is a $d$-path $\pi_i$ of length at most $d$ joining $s_i$ to $s_{i+1}$, and such that $\mathcal C=\pi_0 \cup \ldots \cup \pi_{k-1}$ is a $d$-hull of $\mathcal P$; $s_i \in \mathcal P$, $i=0, \ldots , k-1$, addition taken
\mbox{\rm mod} $k$. 
\end{prob}

By a solution to the MinStation problem we refer simply to a set $S$ of base stations together with the collection of $d$-paths $\pi_i$ whose union is a polygon that encloses $\mathcal P$. 
Recall that a solution is optimal if it contains the least possible number of base stations, and quasi-optimal if it contains one more base station than an optimal solution. 

We also study the next problem, a kind of dual problem to the MinStation problem.
Suppose that we have a budget that allows us to build $k$ base stations, and want to find the locations along $\mathcal P$ where to build them such that
the flight range of the drones used to patrol $\mathcal P$ is minimized, formally:

\begin{prob}[{MinDistance}]
Given a simple polygon $\mathcal P$ and an integer $k\ge 2$, find the smallest $d$ %such that there is
and a set $S=\{s_0,s_2,\dots,s_{k-1}\}$ of $k$ stations on $\mathcal P$ such that, for $i=0,\ldots,k-1$, there is a $d$-path $\pi_i$ of length at most $d$ joining $s_i$ to $s_{i+1}$, addition taken $\modu k$, and $\mathcal C=\pi_0\cup \ldots \cup \pi_{k-1}$ is a $d$-hull of $\mathcal P$.
\end{prob}

\begin{figure}[h!]
    \begin{center}
        \begin{tikzpicture}[scale=0.85]
            %\useasboundingbox (2.35,-1.3) rectangle (9,6.8);
            \draw [color=darkgray] (1.,5.)-- (1.,1.7)-- (4.,0.32)-- (7.,1.7)-- (7.,5.)-- (5.8,3.4)-- (4.6,4.4)-- (4.,3.82)-- (3.4,4.4)-- (2.2,3.4)-- (1.,5.);
            
            \draw [line width=1.2pt,dash pattern=on 4pt off 4pt,color=black] (1.,5.)-- (3.4,4.4)-- (4.,3.82)-- (4.6,4.4)-- (7.,5.)-- (7.,1.7) -- (4.,0.32)-- (1.,1.7) --(1.,5.);
            
            \draw (0.47,3.65) node[anchor=north west] {$d$};
            \draw (6.95,3.65) node[anchor=north west] {$d$};
            \draw (1.6,5.4) node[anchor=north west] {$d - \epsilon$};
            \draw (5.1,5.4) node[anchor=north west] {$d - \epsilon$};
            \draw (3.55,4.57) node[anchor=north west] {$\epsilon$};
            \draw (3.99,4.57) node[anchor=north west] {$\epsilon$};
            \draw (1.98,1.13) node[anchor=north west] {$d$};
            \draw (5.48,1.13) node[anchor=north west] {$d$};
            
            \draw [fill=black] (1.,5.) circle (3pt);
            \draw [fill=black] (1.,1.7) circle (3pt);
            \draw [fill=black] (7.,5.) circle (3pt);
            \draw [fill=black] (7.,1.7) circle (3pt);
            \draw [fill=black] (4.,0.32) circle (3pt);
            \draw [fill=black] (4.,3.82) circle (3pt);
        
            %\draw (current bounding box.south east) rectangle (current bounding box.north west);
        \end{tikzpicture}
    \end{center}
    \caption{Example. The optimal $d$-hull requires 6 base stations. Replacing it with one with smaller perimeter increases the number of base stations to 7.}
    \label{fig:island2}
\end{figure}
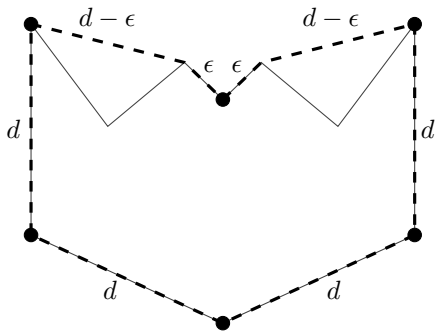

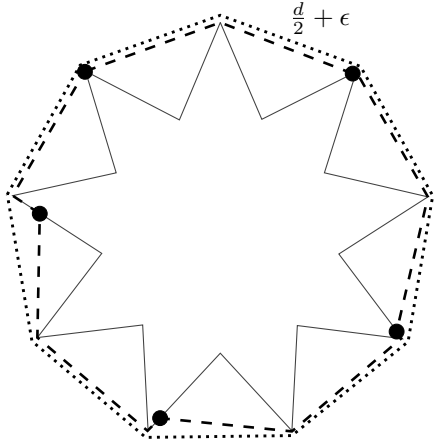
\begin{figure}[htb]
    \begin{center}
        \begin{tikzpicture}[scale=0.8]
        
            \draw [color=darkgray] (4,7.5)-- (4.69,5.89)-- (6.26,6.68)-- (5.73,5)-- (7.45,4.6)-- (5.97,3.65)-- (7.03,2.24)-- (5.28,2.46)-- (5.19,0.71)-- (4,2)-- (2.8,0.71)-- (2.71,2.47)-- (0.96,2.26)-- (2.03,3.66)-- (0.55,4.62)-- (2.27,5)-- (1.76,6.69)-- (3.32,5.88)--(4,7.5);
            
            \draw [line width=1pt,dash pattern=on 4pt off 4pt] (1.76,6.69)-- (4,7.5)-- (6.2,6.65)-- (6.26,6.68)-- (7.45,4.6)-- (6.93,2.36)-- (7,2.24)-- (5.19,0.7)-- (3,0.92)-- (2.79,0.71)-- (0.96,2.26)-- (1,4.32)-- (0.55,4.62)-- (1.75,6.69);
            
            \draw [black, dotted, line width=1.1pt]
            (0.46,4.64)-- (1.67,6.79)-- (4,7.62)-- (6.33,6.77)-- (7.53,4.62)-- (7.12,2.19)-- (5.19,0.63)-- (2.76,0.6)-- (0.87,2.2)-- (0.46,4.64);
            
            \draw (5,8) node[anchor=north west] {$\frac{d}{2} + \epsilon$};
            %\draw (5,8) node[anchor=north west] {$d/2 + \epsilon$};
            
            \draw [fill=black] (1.75,6.68) circle (3.5pt);
            \draw [fill=black] (6.2,6.65) circle (3.5pt);
            \draw [fill=black] (6.93,2.36) circle (3.5pt);
            \draw [fill=black] (3,0.92) circle (3.5pt);
            \draw [fill=black] (1,4.32) circle (3.5pt);
             %\draw (current bounding box.south east) rectangle (current bounding box.north west);
        \end{tikzpicture}
    \end{center}
    \caption{Example. The shortest $d$-hull (dotted) requires almost twice as many stations as the optimal $d$-hull that solves the MinStation problem (dashed). This example can be extended to polygons with arbitrarily many vertices.}
    \label{fig:Star}
\end{figure}

Computing a $d$-hull that minimizes the number of base stations needed to solve the MinStation problem is more subtle than it may at first look.
There are polygons $\mathcal P$ for which, given $d$, the smallest number of base stations needed to solve the MinStation problem, lie on the shortest $d$-hull enclosing $\mathcal P$. 
An example is shown in Figure~\ref{fig:island1}. 
However, there are examples for which the stations of an optimal solution do not lie on the shortest $d$-hull enclosing $\mathcal P$. 
An example is given in Figure~\ref{fig:island2}.
It is easy to see that placing a base station at any point other than the black points shown there, increases the number of base stations needed to solve the MinStation problem. 
In fact, it is not hard to construct polygons such that the number of stations required for the shortest $d$-hull is almost twice the number of stations given by the MinStation problem. 
This is the case, for example, for a star shaped polygon such that the distance between adjacent vertices on the boundary of the convex hull is $\frac{d}{2}+\epsilon$, for some arbitrarily small $\epsilon$, as shown in Figure~\ref{fig:Star}.

We remark that in the optimal solutions of the MinStation and the MinDistance problems, the base stations lie on $\mathcal P$ but not necessarily on vertices of $\mathcal P$ or $\mathcal C$.

\section{Preliminary results} 

Given a fixed point $s \in \mathcal P$ we define a total order $O_s(\mathcal P, \preceq)$ on the points in $\mathcal P$ as follows:
\begin{enumerate}
\item for any point $a \in \mathcal P$, $s \preceq a$
\item for any $a,b \in \mathcal P$, both different from $s$, $a \preceq b$ if $[s,a] \subseteq [s,b]$. (Note that possibly $a=b$.)
\end{enumerate}

For convenience we will add an extra element $s'$ to our order such that for any $a \in \mathcal P$, $a \preceq s'$; that is, $s$ and $s'$ are, respectively, the minimum and the maximum elements of $O_s(\mathcal P, \preceq)$. 
We can think of $s'$ as a copy of $s$, and refer to $O_s(\mathcal P, \preceq)$ simply as $\preceq$.

Consider a point $s \in \mathcal P$ and the order $\preceq$ it defines on the points on $\mathcal P$. 
We define a distance $\delta_d$ on the points on $\mathcal P$ as follows:
\begin{enumerate}
\item $\delta_d(a,a)=0$.
\item If $a \preceq b \in \mathcal P$,  $\delta_d(a,b)=1$ if there is a $d$-path of length at most $d$ from $a$ to $b$.
\item $\delta_d(a,b)=k$ if $k$ is the smallest integer such that there is a sequence of points $p_0=a, \cdots, p_{k}=b$ such that  $\delta_d(p_i, p_{i+1})=1$, $i=0, \ldots, k-1$. 
\end{enumerate}

The following technical Lemma will be crucial in the proposed approach to solve the MinStation problem.

\begin{lemma}[The Sandwich Lemma]
\label{lem:sandwich}
Let $w,x,y,z \in \mathcal P$ such that $w \preceq x\preceq y \preceq z$ on $\mathcal P$, such that $\delta_d(w,y)\leq 1$,  and $\delta_d(x,z)\leq 1$. Then  $\delta_d(w,z)\leq 2$, $\delta_d(w,x)\leq 2$ and $\delta_d(y,z)\leq 2$.
\end{lemma}
\noindent

\begin{proof}
Suppose that $\delta_d(w,z) > 1$, for otherwise we are finished.
Since $w \preceq x\preceq y \preceq z$ the shortest $d$-paths $\pi_{w,y}$ and $\pi_{x,z}$ joining $w$ to $y$, and $x$ to $z$ intersect. 
Let $p$ be a point in the intersection of $\pi_{w,y}$ and $\pi_{x,z}$.
If the distance $\delta(x,p)$ along $\pi_{x,z}$ between $p$ and $x$ is smaller than the distance $\delta(p,y)$ between $p$ and $y$ along $\pi_{w,y}$, then
$\delta(w, p) + \delta(p, x) \leq 1$ and therefore $\delta_d(w, z) \leq 2$. 
The case when $\delta(p, x) \geq \delta(p, y)$ follows the same way. 
The inequalities $\delta_d(w,x)\leq 2$ and $\delta_d(y,z)\leq 2$ are proved in a similar way.
\end{proof}

\begin{figure}[htb]
    \begin{center}
        \begin{tikzpicture}[scale=0.55, rotate=180]
            \useasboundingbox (2.35,-1.3) rectangle (9,6.8);
            
            \draw[color=darkgray] (5,0)--(7,2)--(6,2)--(7,4)--(5,3)--(4,4)--(3,2);
            
            \draw[line width=0.8pt, black, dashed] (5,0)--(5,3);
            \draw[line width=0.8pt, black, dashed] (6,2)--(3,2);
            
            \draw[fill=black] (5,0) circle (3.5pt);
            \draw[fill=black] (6,2) circle (3.5pt);
            \draw[fill=black] (5,3) circle (3.5pt);
            \draw[fill=black] (3,2) circle (3.5pt);
            \draw[fill=black] (5,2) circle (3.5pt);
            
            \draw [color=darkgray] plot [smooth, tension=2] coordinates { (5,0) (7,0) (8,5) (4,5) (3,2)};
            
            \node at (4.5,0) {$w$};
            \node at (6,1.5) {$x$};
            \node at (5,3.5) {$y$};
            \node at (3,1.5) {$z$};
            \node at (4.5,1.5) {$p$};
            %\node at (7,5) {$\mathcal P$};
            
            %\draw (current bounding box.south east) rectangle (current bounding box.north west);
        \end{tikzpicture}
        \caption{Illustration of Lemma~\ref{lem:sandwich}.}
    \end{center}
    \label{f1}
\end{figure}
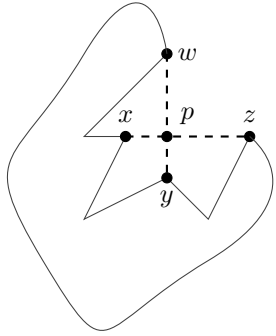

The Sandwich Lemma suggests that in an optimal solution to the MinStation problem, a drone flies around in a non-crossing curve $\mathcal C$ that encloses $\mathcal P$. 
We formalize this observation in the lemma that follows:

\begin{lemma}
\label{lem:s_forward}
Suppose that $s_0\in \mathcal P \cap \CH(\mathcal P)$ and let $S=\lbrace s_0,s_1,...,s_{k-1}\rbrace$ be a solution to the MinStation problem that goes around $P$ in the clockwise direction and which has the least number of stations among all solutions starting from $s_0$. 
Then $s_0 \preceq s_1\preceq s_2\preceq \dots \preceq s_{k-1}$.
\end{lemma}

\begin{proof}
Assume that all of the $\pi_i$ paths joining $s_i$ to $s_{i+1}$ are of minimum length. 
Since $\mathcal C= \pi_0 \cup \cdots \cup \pi_{k-1}$ encloses $\mathcal P$, any point $p$ in $\CH(\mathcal P)$ lies on $\mathcal C$.
It is now easy to see that $\mathcal C$ \emph{covers} $p$ exactly once. 
It follows now that $s_0$ is not in the interior of $\pi_0$, and that $\pi_0$ is a simple curve that always advances in the clockwise direction along $\mathcal C$.

Now, suppose that $s_{i} \preceq s_{i-1}$ for some $i > 1$, and let $j$ be the maximum value such that $s_j \preceq s_i$; that is, $s_j \preceq s_i \preceq s_{j+1}$.
Using Lemma~\ref{lem:sandwich} it follows that $s_j \preceq s_i \preceq s_{j+1} \preceq s_{i-1}$. 
Thus, by Lemma~\ref{lem:sandwich}, $\delta_d(s_j, s_i) \leq 2$, and since $S$ is an optimal solution, it follows that $s_{j+1}=s_{i-1}$.

Let $r$ be the minimum value such that $s_{i-1} \preceq s_r$. 
Then, we have that $s_i \preceq s_{r-1} \preceq s_{i-1} \preceq s_r$. 
It follows that $s_i=s_{r-1}$.

Now, since $s_{j+1}=s_{i-1}$ and $s_i=s_{r-1}$ we have that $s_j \preceq s_i \preceq s_{i-1} \preceq s_r$, where $\delta_d(s_j, s_{i-1}) = \delta_d(s_i, s_r) = 1$.
Therefore, by Lemma~\ref{lem:sandwich}, $\delta_d(s_j, s_r) \leq 2$. 
This is a contradiction, and thus $s_i \preceq s_{i+1}$ for all $i$. 
Hence, $\lbrace s_0,s_1,...,s_r\rbrace$ continues to make forward progress and the result follows. 
\end{proof}

A similar argument shows that for any optimal solution $S=\lbrace s_0,s_1,...,s_{k-1}\rbrace$ of the MinStation Problem (no longer subject to the condition that $s_0$ is fixed) $\mathcal C$ is a simple closed curve.

\section{The algorithm}

We consider the following algorithm, which constructs a solution to the MinStation problem starting at a point $v \in \mathcal P \cap \CH(\mathcal P)$.

\begin{algorithm}%[H]
	\DontPrintSemicolon
	\SetAlgoSkip{medskip}
	\KwIn{Polygon $\P$, $s_0\in\P\cap\CH(\P)$, $d>0$}
	\KwOut{The stations in an optimal or quasi-optimal $d$-hull for $\P$}
	
	\nl{Let $s_0 = s_{0}' = y_{-1} = v$ and $y_0 = \max\{y: \delta_d(s_0,y) = 1 \}$}
	
	\nl{Set $S_0 = \{s_0\}$ and $i=0$}
	
	\nl\Repeat{$y_{i} = s_{0}'$}{
	
		\nl{$i=i+1$}
		
		\nl{$y_{i}=\max\{y:\delta_d(s_{i-1},y)=2\}$}
		
		\nl{$s_{i} = \mbox{any\ } s\in \{w:\delta_d(s_{i-1},w)=1 \mbox{ and } \delta_d(w, y_{i})=1\}$}
		
		\nl{Set $S_i = S_{i-1} \cup \{s_i\}$}
	}
	
	\Return {the last generated set $S = S_i$}
	\caption{\OptSol}
	\label{alg:opt_sol}
\end{algorithm}

We claim that if we further require that $v\in S$, then the set $S$ returned is, indeed, an optimal solution to MinStation. 
On the other hand, we observe that this algorithm always gives a solution that is globally optimal or quasi-optimal (no longer subject to the restriction that $v\in S$).

\begin{theorem}
\label{thm:optimal_convex_hull}
Given a starting point $s_0 \in \mathcal P \cap \CH(\mathcal P)$, if $k$ is the least value such that $s_k=s_{0}'$, then the set of points  $S=\{s_0,\cdots, s_{k-1}\}$ returned by the \OptSol\ algorithm is an optimal solution to the MinStation problem with the additional requirement that a base station be located at $s_0$.
\end{theorem}

\begin{proof}
Suppose that $S$ has more than one element, for otherwise our result is obvious.
Suppose that $Z=\{z_0, \ldots, z_{n-1}\}$ is an optimal solution for the MinStation problem such that there is a base station located at $s_0=z_0$. 
We prove now that $n=k$

By Lemma~\ref{lem:s_forward}, we may assume that $v = z_0 \preceq z_1\preceq \cdots \preceq z_{n-1}$, that for all $i$, $\delta_d(z_i,z_{i+1})=1$, and that $\delta_d(z_{n-1},s_{0}')= 1$.
Consider now the set $S=\{s_0 = v,\cdots,s_{k-1}\}$ returned by the \OptSol\ algorithm.
Recall that $y_{k-1}=s_0'$.
While the relationship between $S$ and $Z$ is unclear, the relationship between $Z$ and $Y_{k-1} = \{y_{-1}, y_0, \ldots , y_{k-1}\}$ is more straightforward; indeed we prove by induction that, for all $i$, $z_{i+1} \preceq y_{i}$.

This clearly holds for $i=0$, as $z_1 \preceq y_0$ by definition.  Now, suppose $z_i \preceq y_{i-1}$.  Let $j$ be minimal so that $z_{i} \preceq y_{j-1}$.  
Then $j \leq i$, and as $v = y_{-1} \prec z_i$ we have that $j \geq 1$.  
Now, by the minimality of $j$, we know that $y_{j-2}\preceq z_{i}\preceq y_{j-1}$ and by definition of $y_{j-2}$, we have that $s_{j-1} \preceq y_{j-2}$.  
Combining, $s_{j-1}\preceq z_i \preceq y_{j-1}$.  
Then as $\delta_d(s_{j-1}, y_{j-1}) =1$ by construction, Lemma~\ref{lem:sandwich} implies that $z_{i+1} \preceq y_{j} \preceq y_{i}$.  
This completes the inductive step, and hence the proof.   

Therefore, the minimal $k$ such that $\delta_d(s_k,s_0')=1$ and the minimal $n$ such that $\delta_d(y_n,s_0')=1$ are the same, and thus $S=\{s_0 = v,\cdots s_{k-1}\}$ is an optimal solution to the MinStation problem, with the additional requirement that there is a base station at $s_0=v$.
\end{proof}

\begin{theorem} 
\label{thm:optimal_plus_one}
Let $s_0 = v \in \mathcal P \cap \CH(\mathcal P)$. 
The set $S=\{s_0, \ldots , s_{k-1}\}$ returned by the \OptSol\ algorithm is an optimal solution or a quasi-optimal solution to MinStation problem. 
\end{theorem}
\begin{proof} 
Suppose that $Z=\{z_0, \ldots , z_{n-1} \}$ is an optimal solution to the MinStation problem, and that $S=\{s_0, \ldots , s_{k-1}\}$ is the solution returned by the MinStation algorithm. 
We prove now that $k=n$ or $k=n+1$.

  Since $s_0$ is on the convex hull of $\mathcal P$ there is a shortest $d$-path between some $z_i$ and $z_{i+1}$ that contains $s_0$. 
  Hence, adding $s_0$ to $Z=\{z_0, \dots, z_{n-1}\}$ yields an optimal or quasi-optimal solution to the MinStation problem including $s_0$.  
\end{proof}

\begin{remark}\rm
The \OptSol\ algorithm can be easily adapted to solve the MinStation problem for a polygonal line $\mathcal{Q}$ with endpoints $q_0$ and $q_{k}$, contained in $\P$.
This may be useful to patrol a section of the coastline instead of a complete island. 
If both endpoints of $\mathcal{Q}$ are contained in an interval $[a, b]$ such that $\overline{a b}$ is a $d$-bridge of $\P$, then we need at most two base stations depending on the length of $\overline{a b}$.
Otherwise, if $q_0$ is not contained in such an interval $[a, b]$, we run \OptSol\ clockwise starting from $q_0$ and stop when we reach a point $s$ such that $q_k \in [q_0, s]$.
In the remaining case we run \OptSol\ counterclockwise starting from $q_k$.
As the solution returned by \OptSol\ for $\P$ is optimal or quasi-optimal, the solution obtained for $\mathcal{Q}$ is also optimal or quasi-optimal.
\end{remark}

\begin{remark}\rm
Although there are polygons such that the optimal solution contains no points in $\CH(\mathcal{P})$ (a simple modification of Figure~\ref{fig:island1} yields one such example), from a practical point of view, it is convenient to assume that at least one station $v$ lies in $P\cap\CH(\mathcal{P})$, as otherwise any solution that includes $v$ could have an arbitrarily large number of stations (imagine that it is located in a large pocket where bridges cannot be established). 
\end{remark}

\subsection{Time complexity}

We prove now that we can implement the \OptSol\ algorithm to run in $O(n^2)$ time.

\begin{figure}[h!]
	\centering
	\begin{subfigure}[t]{0.48\linewidth}
		\begin{tikzpicture}[scale=0.46]
            \useasboundingbox (0.4,2.4) rectangle (12.6,10.7);
            \draw[] (5.96,6.66) -- (6.1,6.98) -- (5.77,7.12) -- (5.64,6.8) -- cycle; 
            
            \draw [black] plot [smooth, tension=.8] coordinates {(3,9) (3.1, 7.2) (3.5,6.5) (4,7.5)};
            
            %\draw [] (4.5,7)-- (9,4.5);
            \draw [] (4,7.5)-- (8, 5.8);
            
            \draw [black] plot [smooth, tension=.5] coordinates { (8, 5.8) (8.4, 4.7) (10,5)  (8.5, 6.5) (9.5, 7.5) (8, 8.5)};
            
           \draw [] (8,8.5)-- (7,10);
            
            \draw [black] plot [smooth, tension=1] coordinates { (7, 10) (12.5,5.5) (7.8,2.7) (1.1,3.8) (1.8,6.3) (1,8.5) (3,9)};
            
            \draw [line width = 0.8pt, black, dashed] (3,9)-- (5.64,6.8);
            \draw [line width = 0.8pt, black, dashed] (5.64,6.8)-- (7, 10);
            
            \draw [fill=black] (3,9) circle (3.5pt);
            \draw [fill=black] (7,10) circle (3.5pt);
            \draw[color=black] (6.48,10.18) node {$p_j$};
            \draw [fill=black] (8, 5.8) circle (3.5pt);            
            \draw [fill=black] (4,7.5) circle (3.5pt);
            \draw[color=black] (7,6.6) node {$e$};
            \draw [fill=black] (7.58, 9.71) circle (3.5pt);
            \draw[color=black] (8.5,9.99) node {$y_{i+1}$};
            \draw [fill=black] (5.64,6.8) circle (3.5pt);
            \draw[color=black] (5.2, 6.4) node {$s_{i+1}$};
            \draw [fill=black] (2.39,9) circle (3.5pt);
            \draw[color=black] (2.2,9.5) node {$s_i$};
            
            %\draw[color=black] (3,4.5) node {$\mathcal{P}$};
            %\draw (current bounding box.south east) rectangle (current bounding box.north west);
        \end{tikzpicture}

		\caption{\label{fig:2-hop_a}}
		
	\end{subfigure}
	~ 
	\begin{subfigure}[t]{0.48\linewidth}
	    \begin{tikzpicture}[scale=0.46]
            \useasboundingbox (0.4,2.4) rectangle (12.6,10.7);
            
            \draw[color= black] (6.54,5.98) -- (6.69,6.27) -- (6.39,6.42) -- (6.24,6.12) -- cycle; 
            
            \draw [color=darkgray] plot [smooth, tension=.8] coordinates {(4.,9.) (3.1, 7.2) (3.5,6.5) (4.5,7.)};
            
            \draw [color=darkgray] (4.5,7.)-- (7.5,5.5);
            
            \draw [color=darkgray] plot [smooth, tension=.75] coordinates {(7.5,5.5) (7.5,4.5) (5.58, 4.42) (5.5, 3.8) (10.5,3.7) (9.5,5.) (11, 5.7)  (10.4,6.6) (9.,6.)};
            
            \draw [color=darkgray] (9.,6.)-- (8.,10.5);
            
            \draw [color=darkgray] plot [smooth, tension=1] coordinates { (8.,10.5) (12.5,5.5) (7.8,2.7) (1.1,3.8) (1.8,6.3) (1.,8.5) (4.,9.)};
            
            \draw [line width=0.8pt, black, dashed] (4.,9.)-- (6.24,6.12);
            \draw [line width=0.8pt, black, dashed] (6.24,6.12)-- (8.13348519068735,9.9);
            
            \draw [fill=black] (4.,9.) circle (3.5pt);
            \draw [fill=black] (8.,10.5) circle (3.5pt);
            
            \draw [fill=black] (9.,6.) circle (3.5pt);
           
           \draw[color=black] (9,8) node {$f$};
           
           \draw[color=black] (6.8,8.2) node {$d$};
            
            \draw [fill=black] (7.5,5.5) circle (3.5pt);
            
            \draw [fill=black] (4.5,7.) circle (3.5pt);
            
            \draw[color=black] (5.1,6.25) node {$e$};
            
            \draw [fill=black] (8.13,9.86) circle (3.5pt);
            \draw[color=black] (7.2,9.85) node {$y_{i+1}$};
            
            \draw [fill=black] (6.25,6.125) circle (3.5pt);
            \draw[color=black] (6.2,5.6) node {$s_{i+1}$};
            
            \draw [fill=black] (3.4,9.05) circle (3.5pt);
            \draw[color=black] (3.5,9.45) node {$s_i$};
            
            %\draw (current bounding box.south east) rectangle (current bounding box.north west);
        \end{tikzpicture}
        \caption{\label{fig:2-hop_b}}
	\end{subfigure}
	
	\caption{The $2$-hop problem. (a) $s_{i+1}$ is a projection of the vertex $p_j$ on the edge $e$. (b) $s_{i+1}$ is a $d$-projection of the edge $f$ on the edge $e$.}
	\label{fig:2-hop}
\end{figure}
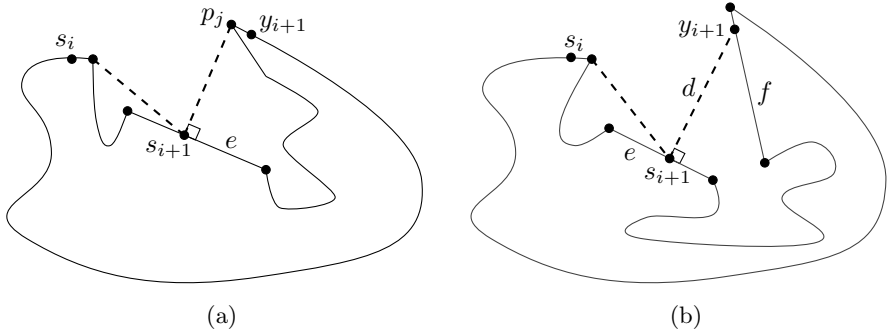

Given a point $s_i$ we want to find a point $y_{i+1} = \max\{y:\delta_d(s_i,y)=2\}$ with respect to $\preceq$ and a point $s_{i+1}\in \{w:\delta_d(s_i,w)=1 \wedge \delta_d(w, y_{i+1})=1\}$.
We refer to the problem of finding $s_{i+1}$ and $y_{i+1}$ as the $2$-\emph{hop} problem, see Figure~\ref{fig:2-hop}.

We will show that by applying a quadratic time pre-processing on $\mathcal P$ the $2$-hop problem can be solved in linear time for each $s_i$. 

A point $x$ of an edge $e$ is a \emph{projection} of a vertex $p_i$ on $e$ if $x \preceq p_i$ and the line segment joining them is a $d$-bridge of $\mathcal P$ perpendicular to $e$. 
See Figure~\ref{fig:2-hop_a}.

In a similar way, we say that a point $x$ of an edge $e$ of $\mathcal{P}$ is called a \emph{$d$-projection} of an edge $f$ on $e$ if there is a point $y \in f$ such that the line segment joining them
is a bridge of $\mathcal P$ of length $d$ perpendicular to $e$. See Figure~\ref{fig:2-hop_b}. 

\begin{lemma}
\label{lem:special_points}
Given $s_i$, $s_{i+1}$ is either a vertex of $\mathcal{P}$, the projection of a vertex on an edge, the $d$-projection of an edge or a point with $\delta(s_i, s_{i+1}) = d$.
\end{lemma}
\begin{proof}
Suppose that $s_{i+1}$ is not a vertex of $\mathcal{P}$ and $\delta(s_i,s_{i+1}) < d$.
Let $e$ be the edge of $\mathcal{P}$ containing $s_{i+1}$, see Figure~\ref{fig:2-hop}.
Note that $\delta(s_{i+1}, y_{i+1})=d$ by the choice of $y_{i+1}$.
If $s_{i+1}$ is neither the projection of a vertex on $e$ nor the $d$-projection of an edge on $e$, then it can be moved slightly along edge $e$ and advance $y_{i+1}$. 
This contradicts the definition of $y_{i+1}$.
\end{proof}

There might be $O(n)$ points at distance $d$ from a previously placed station $s_i$. 
However, we only need to consider the maximum with respect to $\preceq$ among them as a candidate for placing $s_{i+1}$, as we prove next.

\begin{lemma}
\label{lem:distance1}
Let $w, x, y, z$ be points in $\mathcal{P}$ such that $w \preceq x \preceq y \preceq z$. 
Suppose that $\delta(w, x) = \delta(w, y) = d$, $\delta(w, z) > d$, and $\delta(x, z) = \ell$. 
Then, $\delta(y, z) \leq \ell$.
\end{lemma}
\begin{proof}
Let $r$ be an intersection point of the shortest $d$-path $\pi_{w, y}$ from $w$ to $y$ and the shortest $d$-path $\pi_{x, z}$ from $x$ to $z$. 
Note that $r$ always exists by the choice of the four points on $\mathcal P$.
Let $\delta(w, r)$ and $\delta(r, y)$ be the distance along $\pi_{w, y}$ between $w$ and $r$, and between $r$ and $y$, respectively. 
Let $\delta(x, r)$ and $\delta(r, z)$ be the distance along $\pi_{x, z}$ between $x$ and $r$, and  between $r$ and $z$, respectively.
Since $\delta(w, z) > d$, we have $\delta(r, y) < \delta(r, z)$. 
Now suppose that $\delta(r, y) > \delta(x, r)$. 
Then we have that $\delta(w, r) + \delta(r, x) < d$, which is as contradiction to our assumption that $\delta(w, x) = d$. 
Thus, $\delta(r,y) \leq \delta(x, r)$ and $\delta(y, z) \leq \ell$.
\end{proof}

We claim that, even though there might be $O(n^2)$ projections of vertices and $d$-projections of edges, $O(n)$ candidate points are sufficient to compute $s_{i+1}$.

Let $e$ and $f$ be edges of $\mathcal{P}$. We say that $e \lessdot f$ if  for any point $x$ in the interior of $e$ and any point $y$ in the interior of $f$, $x \preceq y$.

\begin{figure}[h!]
	\centering
	
	\begin{subfigure}[t]{0.48\linewidth}
	    
		\begin{tikzpicture}[scale = 0.47]
            \useasboundingbox (0,-7.2) rectangle (11,1.7);
            \draw[] (3.35,-5) -- (3.35,-4.65) -- (3,-4.65) -- (3,-5) -- cycle; 
            \draw[] (6.34,-5) -- (6.34,-4.65) -- (6,-4.65) -- (6,-5.) -- cycle; 
            
            \draw [color=darkgray] (1,-5)-- (9.,-5);
            
            \draw [color=darkgray] (9,-0.8)-- (4.5,0.5)--(1,-0.5);
            
            \draw [color=darkgray] plot [smooth, tension=1.4] coordinates { (9,-5.) (9.5,-2.7) (9,-0.8)};
            
            \draw [color=darkgray] plot [smooth, tension=1] coordinates { (1,-0.5) (.5, .5) (6, 1.3) (10, 0.7) (10.5, -2.5) (8.3, -6.5) (3, -6.5) (1, -5)};
            
            \draw [line width=0.8pt, black, dashed] (3,-5)-- (3,0.07);
            \draw [line width=0.8pt, black, dashed] (6,-5)-- (6,0.06);

            \draw (4.3,-4.22) node[anchor=north west] {$e$};
            \draw (1.65,0.85) node[anchor=north west] {$f$};
            \draw (6.64,0.88) node[anchor=north west] {$f'$};
            %\draw (9,0.6) node[anchor=north west] {$\mathcal{P}$};
            
            \draw [fill=black] (6,0.06) circle (3.5pt);
            \draw[color=black] (5.55,-0.32) node {$y'$};
            \draw [fill=black] (3,0.07) circle (3.5pt);
            \draw[color=black] (3.5,-0.36) node {$y$};
            \draw [fill=black] (6,-5.) circle (3.5pt);
            \draw[color=black] (6.15,-5.45) node {$x'$};
            \draw [fill=black] (3,-5) circle (3.5pt);
            \draw[color=black] (3.125,-5.52) node {$x$};
            %\draw (current bounding box.south east) rectangle (current bounding box.north west);
        \end{tikzpicture}
        \caption{}
		\label{fig:edge_projection_1}
	\end{subfigure}
	~
	\begin{subfigure}[t]{0.48\linewidth}
		\begin{tikzpicture}[scale = 0.47]
             \useasboundingbox (0,-2.1) rectangle (11.2,6.8);
            \draw[] (2.92,5) -- (2.92,4.59) -- (3.33,4.59) -- (3.33,5) -- cycle; 
            \draw[] (5.26,5) -- (5.26,4.59) -- (5.66,4.59) -- (5.66,5) -- cycle; 
            
            \draw [color=darkgray] (1,5)--(8.5,5);
            \draw [color=darkgray] (8,-1)--(4.5,0.5)--(1,-1);

            \draw [color=darkgray] plot [smooth, tension=1] coordinates {(8.5,5) (9.5,1.5) (8,-1)};
            
            \draw [color=darkgray] plot [smooth, tension=1] coordinates {(1, 5) (0.5, 6) (5, 6.5) (10, 5.7) (11, 2) (9.3, -1) (5, -2) (1, -1)};
            
            \draw [line width=0.8pt, black, dashed] (3.33,5)-- (3.33,0);

            \draw [line width=0.8pt, black, dashed] (5.66,5)-- (5.66,0);

            %\draw (9,5.7) node[anchor=north west] {$\mathcal P$};
            \draw (4.0,5) node[anchor=north west] {$e$};
            \draw (1.7,0.56) node[anchor=north west] {$f$};
            \draw (6.55,0.54) node[anchor=north west] {$f'$};
            
            \draw [fill=black] (5.66,0) circle (3.5pt);
            \draw[color=black] (5.71,-0.52) node {$x'$};
            \draw [fill=black] (3.33,0) circle (3.5pt);
            \draw[color=black] (3.30,-0.58) node {$x$};
            \draw [fill=black] (5.66,5) circle (3.5pt);
            \draw[color=black] (5.82,5.66) node {$y'$};
            \draw [fill=black] (3.33,5) circle (3.5pt);
            \draw[color=black] (3.40,5.55) node {$y$};
           %\draw (current bounding box.south east) rectangle (current bounding box.north west);
        \end{tikzpicture}
		\caption{}
		\label{fig:edge_projection_2}
	\end{subfigure}
	
	\caption{(a) Case 1: we only need to store the point $x$ on edge $e$. (b) Case 2: we only need to store the point $x$ on edge $f$.}
	\label{fig:edge_projection}
\end{figure}
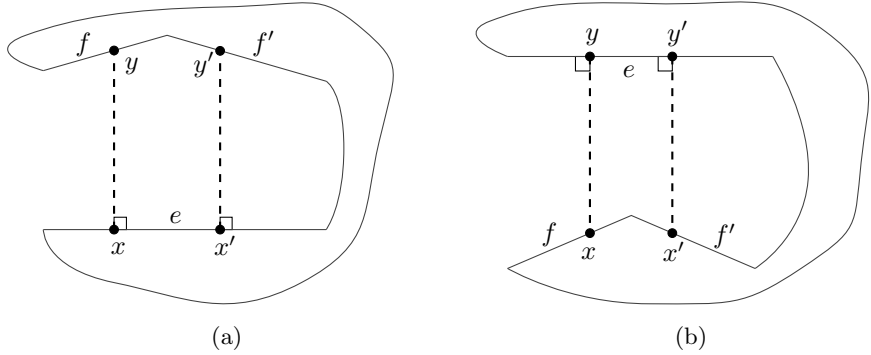

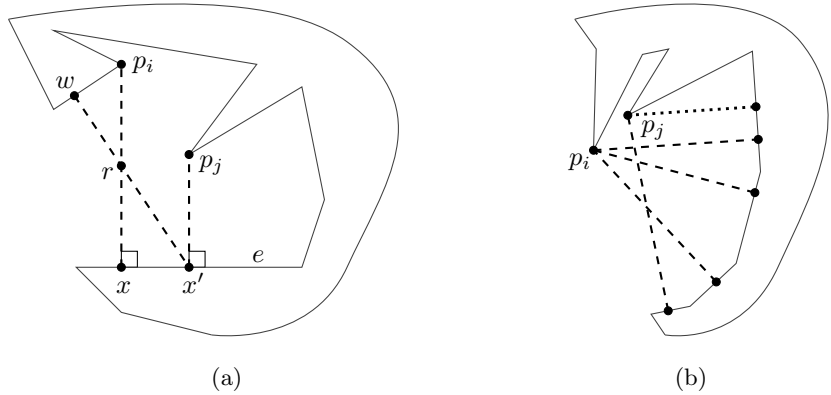
\begin{figure}[ht!]
	\centering
	\begin{subfigure}[t]{0.48\textwidth}
		%\hspace{-10pt}
		\begin{tikzpicture}[scale=0.3]
            \draw[] (13.71,2) -- (13.71,2.71) -- (13,2.71) -- (13,2) -- cycle; 
            
            \draw[] (10.71,2) -- (10.71,2.71) -- (10,2.71) -- (10,2) -- cycle; 
            
            \draw [color=darkgray] (14,-1)-- (10,0)--(8, 2)--(18,2)--(19,5)-- (18,10)-- (13,7)-- (16,11)-- (7,12.5)-- (10,11)-- (7,9)-- (5,13);
            
            \draw [color=darkgray] plot [smooth, tension=1] coordinates {(5, 13) (20, 12) (20, 2) (14, -1)};
            
            \draw [line width = 0.8pt, black, dashed] (7.92,9.61)-- (13,2);
            \draw [line width = 0.8pt, black, dashed] (10,2)-- (10,11);
            \draw [line width = 0.8pt, black, dashed] (13,2)-- (13,7);
            %\draw (17,12) node[anchor=north west] {$\mathcal{P}$};
         
           % \draw [fill=black] (19.,5.) circle (3.0pt);
            \draw [fill=black] (13,7) circle (5pt);
            \draw[color=black] (14,6.51) node {$p_j$};
            \draw [fill=black] (10,11) circle (5pt);
            \draw[color=black] (11,10.98) node {$p_i$};
            \draw[color=black] (16.07,2.5) node {$e$};
            \draw [fill=black] (10,2) circle (5pt);
            \draw[color=black] (10.1,1.13) node {$x$};
            \draw [fill=black] (13,2) circle (5pt);
            \draw[color=black] (13.2,1.3) node {$x'$};
            \draw [fill=black] (7.92,9.61) circle (5pt);
            \draw[color=black] (7.51,10.22) node {$w$};
            \draw [fill=black] (10,6.5) circle (5pt);
            \draw[color=black] (9.4,6.16) node {$r$};
        
        \end{tikzpicture}
		\caption{}
		\label{fig:vertex_projection_1}
	\end{subfigure}
	~ 
	\begin{subfigure}[t]{0.48\textwidth}
		\hspace{30pt}
		\begin{tikzpicture}[scale=0.3]
    
            \draw [color=darkgray] (16.,-1.)-- (15.38,-0.08)-- (17.11,0.27)-- (19.15,2.17)-- (20.23,6.24)-- (19.87,11.58)--(14.35,8.74)-- (16.16,11.68)-- (15,11.44)-- (12.83,7.19)-- (12.95,11.68)-- (12.,13.);
            
            \draw [color=darkgray] plot [smooth, tension=1] coordinates {(12, 13) (22, 12) (21, 2) (16, -1)};
            
            \draw [line width = 0.8pt, black, dashed] (19.98,5.31)-- (12.83,7.19);
            
            \draw [line width = 0.8pt, black, dashed] (12.83,7.19)-- (20.13,7.67);
            
            \draw [line width = 1.1pt, black, dotted] (14.35,8.74)-- (20.03,9.12);
            
            \draw [line width = 0.8pt, black, dashed] (18.27,1.35)-- (12.83,7.19);
            \draw [line width = 0.8pt, black, dashed] (16.14,0.074)-- (14.35,8.74);
            
            %\draw (20.1,12.08) node[anchor=north west] {$\mathcal{P}$};
            
            \draw [fill=black] (14.35,8.74) circle (5pt);
            \draw[color=black] (15.5,8.1) node {$p_j$};
            \draw [fill=black] (12.83,7.19) circle (5pt);
            \draw[color=black] (12.24,6.69) node {$p_i$};
            \draw [fill=black] (19.98,5.31) circle (5pt);
            \draw [fill=black] (20.13,7.67) circle (5pt);
            \draw [fill=black] (20.03,9.12) circle (5pt);
            \draw [fill=black] (18.27,1.35) circle (5pt);
            \draw [fill=black] (16.14,0.07) circle (5pt);
        
        \end{tikzpicture}
		\caption{}
		\label{fig:vertex_projection_2}
	\end{subfigure}
	
	\caption{(a) Case 3: The distance from $x$ to $w$ is smaller than the distance from $x'$ to $w$. (b) We need to store all vertex projections except the one that is the endpoint of the dotted segment. }
	\label{fig:vertex_projection}
\end{figure}

\begin{lemma}
\label{lem:projections_1}
For each edge $e$ of $\mathcal{P}$ we need to store at most three points:
\begin{enumerate}
	\item The minimum $d$-projection (with respect to $\preceq$) of an edge $e'$ on $e$ such that $e \lessdot e'$.
	
	\item The endpoint not in $e$ of the bridge generating the maximum $d$-projection (with respect to $\preceq$) on $e$ of an edge $e'$ such that $e' \lessdot e$. In this case the stored point lies on $e'$.
	
	\item The minimum projection (with respect to $\preceq$) of a vertex on $e$.
\end{enumerate}

\end{lemma} 
\begin{proof}
\noindent\textit{Case 1.} 
Let $x$ and $x'$ be $d$-projections of two distinct edges $f$ and $f'$, respectively, on $e$ such that $e \lessdot f$ and $e \lessdot f'$. 
Let $\overline{x  y}$ be the $d$-bridge perpendicular to $e$ having $x$ as an endpoint, i.e., $y \in f$ and $\overline{x y}$ is has length $d$. 
Define $\overline{x' y'}$ analogously.
Because of the length of $\overline{x y}$ (respectively, $\overline{x' y'}$), if we place a station at $x$ (respectively, $x'$) then we also need to place a station at $y$ (respectively, $y'$).
Suppose w.l.o.g. that $x \preceq x'$, see Figure~\ref{fig:edge_projection_1}. 
Since all the bridges defining $d$-projections of edges on $e$ are parallel, this implies that $f' \lessdot f$ and $y' \preceq y$. 
Moreover, as the interval $[x, y]$ contains the interval $[x', y']$, placing a station at $x$ guarantees that both intervals of $\mathcal{P}$ are guarded.
Hence, we maximize $y_{i+1}$ with respect to $\preceq$ by choosing the minimum $d$-projection of an edge on $e$ as $s_{i+1}$. 
\\

\noindent\textit{Case 2.} 
This case is analogous to the first one, see Figure~\ref{fig:edge_projection_2}.
\\

\noindent\textit{Case 3.} 
Let $x$ and $x'$ be the projections of two distinct vertices $p_i$ and $p_j$, respectively, on an edge $e$.
Let $\overline{x p_i}$ and $\overline{x' p_j}$ be their corresponding $d$-bridges. 
Suppose w.l.o.g. that $x \preceq x'$. This implies that $p_j \preceq p_i$ and that placing a station at $x$ guarantees that both intervals $[x, p_i]$ and $[x', p_j]$ are guarded, see Figure~\ref{fig:vertex_projection_1}.
It remains to be proven that by placing a station at $x$ we can advance further on $\mathcal{P}$ with respect to $\preceq$.
Let $w \in \mathcal P$ be a point such that $p_i \preceq w$, and let $p_{x, w}$ and $p_{x', w}$ be the shortest $d$-paths joining $x$ to $w$ and $x'$ to $w$. 
Let $r$ be the intersection point of $\overline{x p_i}$ and $p_{x', w}$.
Notice that the points $x$, $x'$ and $r$ form a right triangle that is right-angled at $x$.
Therefore, the length of $p_{x, w}$ is smaller than the length of $p_{x', w}$, which implies that we can maximize $y_{i+1}$ by choosing the minimum projection of a vertex on $e$ as $s_{i+1}$.

\end{proof}

In order to compute the candidate points on $\mathcal{P}$, we first find, for each edge $e \in P$, the subset containing each point $x \in \mathcal{P}$ for which there is a segment perpendicular to $e$ joining $x$ and $e$, and completely contained in $Ext(\mathcal{P})$.
In such case we say that $x$ is \emph{orthogonally visible} from $e$.

We define a \emph{lid} as an edge of the convex hull of $\mathcal{P}$ that is not an edge of $\mathcal{P}$.
Each lid $h = \overline{a b}$ defines a polygon $\mathcal{P}_h$, which is the union of $h$ and the interval of $\mathcal{P}$ determined by $a$ and $b$ which has no points in the convex hull of $\mathcal{P}$ besides $a$ and $b$. 
Note that any projection of a vertex or $d$-projection of an edge is defined by a segment whose endpoints are contained in the same $\mathcal{P}_h$, for otherwise the segment would intersect $Int(\mathcal{P})$.
Therefore, we only need to compute the set of points orthogonally visible from each edge $e$ contained in a $\mathcal{P}_h$; moreover, we only need to look at the polygon $\mathcal{P}_h$ containing $e$ to find these points.

For the next lemma, we assume that we have computed the polygons defined by all the lids of $\mathcal{P}$, as well as the triangulation of each such polygon. This can be done in $O(n)$ time overall, see \cite{melkman1987line} and \cite{chazelle1991triangulating}.

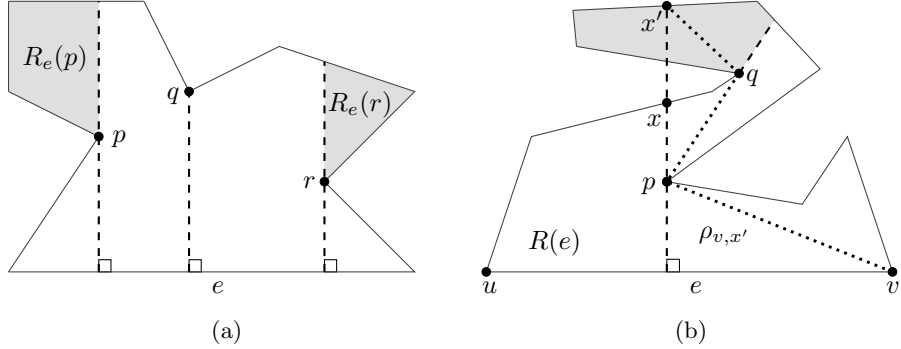
\begin{figure}[h!]
	\centering
	\begin{subfigure}[t]{0.48\linewidth}
	    %\hspace{-25pt}
		\begin{tikzpicture}[scale=0.6]

            \draw[] (7.27,1.) -- (7.27,1.27) -- (7.,1.27) -- (7.,1.) -- cycle; 
            
            \draw[] (4.27,1.) -- (4.27,1.27) -- (4.,1.27) -- (4.,1.) -- cycle; 
            
            \draw[] (2.27,1.) -- (2.27,1.27) -- (2.,1.27) -- (2.,1.) -- cycle;
            
            \fill[fill=gray!25,fill opacity=1] (2.,4.) -- (2,7.) -- (0.,7.) -- (0.,5.) -- cycle;
            
            \fill[fill=gray!25,fill opacity=1] (7.,3.) -- (7,5.66) -- (9.,5.) -- cycle;
            
            \draw [color=darkgray] (0.,1.)-- (9.,1.)-- (7.,3.)-- (9.,5.)-- (6.,6.)-- (4.,5.);
            
            \draw (3.3,5.3) node[anchor=north west] {$\Huge{q}$};
            
            \draw (2.1,4.33) node[anchor=north west] {$\Huge{p}$};
            \draw (6.32,3.33) node[anchor=north west] {$\Huge{r}$};
            
            \draw [color=darkgray] (4.,5.)-- (3.,7.);
            \draw [color=darkgray] (3.,7.)-- (0.,7.);
            \draw [color=darkgray] (2.,4.)-- (0.,1.);
            
            \draw [line width = 0.8pt, black, dashed] (2.,1.)-- (2,7.);
            \draw [line width = 0.8pt, black, dashed] (4.,1.)-- (4.,5.);
            \draw [line width = 0.8pt, black, dashed] (7.,1.)-- (7,5.66);
            
            \draw (0.1,6.24) node[anchor=north west] {$\Huge{R_e(p)}$};
            \draw (4.3,1) node[anchor=north west] {$\Huge{e}$};
            
            \draw [color=darkgray] (0.,7.)-- (0.,5.);
            \draw [color=darkgray] (0.,5.)-- (2.,4.);
            \draw (6.87,5.22) node[anchor=north west] {$\Huge{R_e(r)}$};
            
            \draw [fill=black] (7.,3.) circle (2.7pt);
            \draw [fill=black] (4.,5.) circle (2.7pt);
            \draw [fill=black] (2.,4.) circle (2.7pt);
        \end{tikzpicture}
        \caption{}
		\label{fig:turn_vertices}
		
	\end{subfigure}
	~ 
	\begin{subfigure}[t]{0.48\linewidth}
		%\hspace{20pt}
	    \begin{tikzpicture}[scale=0.6]

            \draw[] (4.28,1) -- (4.28,1.28) -- (4,1.28) -- (4,1) -- cycle; 
            
            \fill[fill=gray!25,fill opacity=1] (5.6,5.4) -- (6.4,6.6) -- (6,7) -- (1.92,6.8) -- (2,6) -- cycle;
            
            \draw [color=darkgray] (0,1)-- (9,1)-- (8,4)-- (7,2.5)-- (4,3)-- (7.4,5.5);
            
            \draw (3.22,7) node[anchor=north west] {$\Huge{x'}$};
            \draw (3.35,4.7) node[anchor=north west] {$\Huge{x}$};
            \draw (5.55,5.7) node[anchor=north west] {$\Huge{q}$};
            
            \draw [color=darkgray] (7.4,5.5)-- (6,7);
            
            \draw [color=darkgray] (6,7)-- (1.92,6.8);
            
            \draw (3.25,3.3) node[anchor=north west] {$\Huge{p}$};
            \draw (4.3,1) node[anchor=north west] {$\Huge{e}$};
            
            \draw [color=darkgray] (1.92,6.8)-- (2,6);
            
            \draw (4.5,2.2) node[anchor=north west] {$\Huge{\rho_{v, x'}}$};
            
            \draw [color=darkgray] (2,6)-- (5.6,5.4)-- (5,5)-- (1,4)-- (0,1);
            
            \draw [line width = 0.8pt, black, dashed] (5.6,5.4)-- (6.38,6.58);
            
            \draw [line width = 0.8pt, black, dashed] (4,1)-- (4,6.9);
            
            \draw [black, dotted, line width=1.1pt] (5.6,5.4)-- (4,3);
            
            \draw (-0.3,1) node[anchor=north west] {$\Huge{u}$};
            \draw (8.65,1) node[anchor=north west] {$\Huge{v}$};
            
            \draw [black, dotted, line width=1.1pt] (9,1)-- (4,3);
            \draw [black, dotted, line width=1.1pt] (5.6,5.4)-- (4,6.9);
            
            \draw (0.72,2.16) node[anchor=north west] {$\Huge{R(e)}$};
            
            \draw [fill=black] (0,1) circle (2.7pt);
            \draw [fill=black] (9,1) circle (2.7pt);
            \draw [fill=black] (4,3) circle (2.7pt);
            \draw [fill=black] (5.6,5.4) circle (2.7pt);
            \draw [fill=black] (4,6.9) circle (2.7pt);
            \draw [fill=black] (4,4.75) circle (2.7pt);
            \draw [fill=black] (4,3) circle (2.7pt);
        \end{tikzpicture}
        \caption{}
        \label{fig:turn_subpolygon}
        
	\end{subfigure}

	\caption{(a) $p$ is a left turn vertex, $r$ is a right turn vertex, and $q$ is not a turn vertex. 
	(b) Neither $x'$ nor any point in the shaded region is in $R_e$: the shortest path from $v$ to $x'$ makes a left turn at $q$. }
	\label{fig:turn}
\end{figure}

\begin{lemma}
\label{lem:visible_segments}
We can find the set containing all the segments of $\mathcal{P}$ orthogonally visible from any edge of $\mathcal{P}$ in $O(n)$ time.
Moreover, each such set has $O(n)$ size.
\end{lemma} 
\begin{proof}
Let $h = \overline{a b}$ be a lid of $\mathcal{P}$ and let $e = \overline{u v}$ be an edge of $\mathcal{P}_h$. We proceed as follows:
Compute the set $V\!P(\mathcal{P}_h, e)$ of points of $\mathcal{P}_h$ visible from a point in $e$. $V\!P(\mathcal{P}_h, e)$ can be computed in $O(n)$ time~\cite{guibas1987linear}.

Suppose w.l.o.g. that $u \prec v$. 
Let $R$ be the region contained between the lines perpendicular to $e$ through $u$ and $v$, and to the left of the line directed from $u$ to $v$.
It is easy to see that any point of $\mathcal{P}$ orthogonally visible from $e$ must lie in $\mathcal{R}_e = V\!P(\mathcal{P}_h, e) \cap R$, which can be computed in $O(n)$ time by intersecting $V\!P(\mathcal{P}_h, e)$ with both lines.
We suppose w.l.o.g. that $e$ is horizontal and that the interior of $\mathcal{R}_e$ lies above $e$.

We say that a vertex $p \in \mathcal{R}_e$ is a \emph{turn vertex} if the maximal vertical segment $\overline{x y}$ through $p$ and completely contained in $\mathcal{R}_e$ separates $\mathcal{R}_e$ into three subpolygons, see Figure~\ref{fig:turn_vertices}. 
If two of these subpolygons lie to the right (left) of $\overline{x y}$, we say that $p$ is a \emph{right (left)} turn vertex.
Let $x$ be the top endpoint of $\overline{x  y}$. 
The segment $\overline{p x}$ separates $\mathcal{R}_e$ into two subpoygons, one of them containing $e$. 
Let $R_e(p)$ denote the subpolygon generated by $\overline{p x}$ not containing $e$.
It is easy to see that any point in $\mathcal{R}_e$ not being orthogonally visible from $e$ lies in the subpolygon $\mathcal{R}_e(p)$ for some turn vertex $p$, and that any point in $\mathcal{R}_e(p) \setminus \overline{px}$ is not orthogonally visible from $e$.

Note that the internal angles at both vertices of $e = \overline{uv}$ are convex in $\mathcal{R}_e$. 
Ghosh et al.~\cite{ghosh1993characterizing} proved that for any vertex $p$ in $\mathcal{R}_e$, the shortest path from $u$ to $p$, denoted as $\rho_{u, p}$, makes a left turn at every vertex of the path, and $\rho_{v, p}$ makes a right turn at every vertex of the path.
This also holds true for the points in the interior of any edge of $\mathcal{R}_e$.

Let $p$ be a turn vertex of $\mathcal{R}_e$ and let $x$ be the top endpoint of the maximal vertical segment through $p$ completely contained in $\mathcal{R}_e$.
We claim that the vertical line through $p$, $\ell_p$, does not intersect any point of $\mathcal{R}_e(p) \setminus \overline{p x}$. 
Suppose otherwise that there is a point $x'$ in $\mathcal{R}_e(p) \setminus \overline{p x}$ contained in $\ell_p$.
Then, there exists a vertex $q$ in $\mathcal{R}_e(p) \setminus \overline{p x}$ such that $\rho_{v, x'}$ makes a left turn at $q$ or $\rho_{u, x'}$ makes a right turn at $q$, which is a contradiction \cite{ghosh1993characterizing}, see Figure~\ref{fig:turn_subpolygon}. 
It follows that $\mathcal{R}_e(p) \cap \ell_p = \overline{p x}$.
This fact yields the following algorithm for removing $\mathcal{R}_e(p)$ from $\mathcal{R}_e$ for each turn vertex $p$.

We deal with the right turn vertices by traversing the edges of $\mathcal{R}_e$ clockwise from $v$ to $u$.
We set a variable \id{edgeIsVisible} to \textit{true}.
Let $f = \overline{q r}$, $q \prec r$, be the current edge in the traversal.
\begin{itemize}
    \item If \id{edgeIsVisible} is \textit{true} we check if $r$ is a       right turn vertex. 
    In the affirmative case, we set \id{edgeIsVisible} to \textit{false} and store the vertical line through $r$, $\ell_r$ and the edge $f$.
          
    \item If \id{edgeIsVisible} is \textit{false}, then we had             previously stored the last visible edge $g = \overline{o p}$,          where $p$ is a right turn vertex, and the vertical segment through $p$, $\ell_p$. 
    We check if $x = f \cap \ell_p$ is not empty. 
    In such a case, we replace the interval $[p, x]$ of $\mathcal{R}_e$ with the vertical segment $\overline{p x}$, set \id{edgeIsVisible} to \textit{true}, and discard $g$ and $\ell_p$.
\end{itemize}
We can remove the sub-polygons defined by the left turn vertices analogously by traversing $\mathcal{R}_e$ counter-clockwise from $u$ to $v$.
As each edge of $\mathcal{R}_e$ is visited at most twice, the removal of the sub-polygons defined by all the turn vertices takes $O(n)$ time.
Let $\mathcal{R}'_e$ be the polygon obtained by these traversals.

To obtain the subset of $\mathcal{P}$ orthogonally visible from $e$, we only need to discard $e$, the segment contained in the lid of $\mathcal{P}_h$, and the vertical segments added in the previous process (at most one per turn vertex) from $\mathcal{R}'_e$.

Since $\mathcal{P}_h$ has no holes, each edge of $\mathcal{P}_h$ provides at most one segment to $\mathcal{R}'_e$.
Therefore, the set of segments of $\mathcal{P}$ orthogonally visible from any edge $e$ has $O(n)$ size.
\end{proof}

\begin{lemma}
\label{lem:projections_2}
For any edge $e$ of $\mathcal{P}$ we can find the projections described in Lemma~\ref{lem:projections_1} in $O(n)$ time.
\end{lemma}
\begin{proof}
Suppose that $e = \overline{u v}$, $u \preceq v$. By Lemma~\ref{lem:visible_segments}, we can find the set $W$ of all the segments of edges and vertices of $\mathcal{P}$ orthogonally visible from $e$ in $O(n)$ time; moreover, $W$ has $O(n)$ size.
Let $W_B$ be the subset of elements of $W$ smaller than $u$ and let $W_A$ be the subset of the elements of $W$ greater than $v$ with respect to $\preceq$.

We find the $d$-projections corresponding to the first two cases of Lemma~\ref{lem:projections_1} as follows.
Let $\ell$ be the line parallel to $e$, to the left of the line directed from $u$ to $v$ and at distance $d$ from $e$.
We first compute the intersection of $\ell$ with both $W_B$ and $W_A$, which by the size of $W$ can be obtained in $O(n)$ time.
To obtain the point described in the first case of the proof of Lemma~\ref{lem:projections_1} we take the maximum point $q$ with respect to $\preceq$ in $\ell \cap W_A$ and store the intersection point of $e$ with the line through $q$ perpendicular to $e$.
To obtain the described in the second case of the proof of Lemma~\ref{lem:projections_1} we store the minimum point in $\ell \cap W_B$ with respect to $\preceq$, if any.

We find the projection of the maximum vertex on $e$ described in the third case of Lemma~\ref{lem:projections_1} as follows.
For each vertex of $\mathcal{P}$ in $W_B$ we compute its distance with respect to $e$. We then store the maximum with respect to $\preceq$ of the vertices at distance less or equal than $d$ from $e$.
\end{proof}

Now we need to solve the following subproblem: given a point $x \in \mathcal{P}$, find the maximum $w$, $x \preceq w$, such that $\delta(x, w) = d$.
Guibas et al.~\cite{guibas1987linear} proved that, given the triangulation of a polygon $\mathcal{R}$ and a point $p \in \mathcal{R}$, the euclidean shortest paths from $p$ to all the vertices of $\mathcal{R}$ can be found in linear time (see also \cite{lee1984euclidean}). 
The union of all the shortest paths from the source point $p$ to the vertices of $\mathcal{R}$ is a planar tree called the \emph{shortest-path tree} of $\mathcal{R}$ with respect to $p$.

Let $\mathcal{R}$ be the polygon obtained by enclosing $\mathcal{P}$ in a sufficiently large rectangle and connecting one of the sides of the rectangle to the starting point of the sequence, $x_0$, with a thin corridor.
The polygon $\mathcal{R}$ can be obtained in $O(n)$ time, see~\cite{ntafos1994external}.
Note that $\mathcal{R}$ has $m \leq n + 8$ vertices and $\mathcal{P}$ is contained in the exterior of $\mathcal R$. %Let $R = B(P, x_0)$.
We assign to the points in $\mathcal{R}$ that are also points in $\mathcal{P}$ the same order as in $\mathcal P$.

Henceforth we assume that $\mathcal{R}$ has been computed along with its triangulation, which as proven by Chazelle~\cite{chazelle1991triangulating} can be found in $O(n)$ time.

\begin{lemma}
\label{lem:max_distance1}
Given any point $x \in \mathcal{P}$, the point $w \in \mathcal{P}$ with $\delta(x, w) = d$ such that $w$ is maximum with respect to $\preceq$ can be found in $O(n)$ time.
\end{lemma}
\begin{proof}
Let $x$ be a point in $\mathcal{P}$ and let $x'$ be its corresponding point in $\mathcal{R}$.
We compute the shortest path $\rho(x', y)$ from $x'$ to every vertex $y \in R$ such that $y$ is also a vertex of $P$ and $x' \prec y$.
Let $T$ be the shortest-path tree obtained by the union of these shortest paths.
Let $M$ be the set of vertices of $T$ such that, for any $w \in M$, $\delta(x', w) \leq d$, and $w$ shares an edge of $\mathcal{R}$ with a vertex $y$ such that $\delta(x', y) > d$.
The set $M$ can be found in $O(n)$ time by traversing $T$ from its root $x'$.

Observe that any point of $\mathcal{R}$ at distance $d$ from $x'$ is one of the following:
\begin{itemize}
	\item An element of $M$.
	
	\item A point in an edge $e = \overline{u v}$, $u \prec v$, of $\mathcal{R}$ such that $e \in E(T)$. 
	In this case, $u \in M$ and $\delta(x', v) > d$.
	
	\item A point in an edge $e = \overline{u v}$, $u \prec v$, of $\mathcal{R}$ such that $e \notin E(T)$. 
	Notice that, in this case, $\delta(x', v) > d$. 
	Moreover, there is exactly one $z \in M$ such that $(z, u) \in E(T)$.
\end{itemize}

Hence, in order to find all the points at distance exactly $d$ from $x'$ it is sufficient to check the edges having a neighbour of an element of $M$ in $T$ as an endpoint.
Since each vertex is adjacent to at most one element of $M$, this can be done in $O(n)$ time.
At the final step we need to find the maximum among all the points at distance $d$ from $x'$, which can also be done in $O(n)$ time.
Our result follows.
\end{proof}

\begin{theorem}
Let $\mathcal{P}$ be a polygon with $n$ vertices and let $s_0 \in \mathcal{P}$ be a point on the convex hull of $\mathcal{P}$. 
Then \OptSol\ returns an optimal solution $S$ to the MinStation Problem such that $s_0 \in S$ in $O(n^2 + \frac{L}{d} n)$ time, where $L$ is the length of~$\mathcal{P}$.
\end{theorem}
\begin{proof}
By Lemma~\ref{lem:special_points}, given $s_i$, the point $s_{i+1}$ is either a point on $\mathcal{P}$ at distance exactly $d$ from $s_i$, a vertex of $\mathcal{P}$, the projection of a vertex onto an edge, or the $d$-projection of an edge onto another edge.

By Lemma~\ref{lem:distance1}, we only need to consider the maximum point with respect to $\preceq$ at distance $d$ from $x_i$, which can be found in $O(n)$ time as stated in Lemma~\ref{lem:max_distance1}.

There might be $O(n^2)$ projections of vertices and $d$-projections of edges. 
However, Lemma~\ref{lem:projections_1} states that in the set of candidates we need to store at most three projections for each edge of $\mathcal{P}$.
Moreover, these projections can be found in $O(n)$ time for each edge.

The set of candidate points to compute all the elements of the set $S$ has $O(n)$ size.
For each candidate $x$, we compute the maximum point at distance $d$ from $x$ and associate this point to $x$, which by Lemma~\ref{lem:max_distance1} takes $O(n)$ time per candidate.

It is easy to see that we only need to consider the candidates contained in the interval of $\mathcal P$ from $s_i$ to the maximum point with respect to $\preceq$ at distance $d$ from $s_i$.
From all these candidates, we choose as $s_{i+1}$ the candidate which maximizes $y_{i+1}$, which can be done in $O(n)$ time.
Since we might need to place $O(\frac{L}{d})$ stations, this step takes time $O(\frac{L}{d} n)$.
Therefore, the set $S$ can be found in $O(n^2 + \frac{L}{d} n)$ time.
\end{proof}

\section{Discretization}

In this section, we present a discretization algorithm that is easy to implement for the MinStation problem, and then show how it can be utilized to obtain a solution to the MinDistance problem which is close to optimal. 
This algorithm avoids computing projections, drone distances (geodesic paths) and  orthogonal visibility, which makes it very practical. 
The idea is to construct a graph and apply a slight modification of Dijkstra algorithm.

Fix $0 < \epsilon \leq d$ and let $X=\{s_0=x_0 \preceq \ldots \preceq x_{r-1}\} \subseteq \mathcal P$ be a set of points so that $s_0$ lies on $\mathcal P \cap \CH(\mathcal P)$ and the distance between $x_i$ and $x_{i+1}$ along $\mathcal P$ is at most $\epsilon$, addition taken $\modr r$. 
For technical reasons that will become apparent later, we also ask that the vertices of $\mathcal{P}$ are contained in $X$. 
Consider the graph $G_d(X)$ such that $V(G_d(X)) = X$ in which two elements $x_i, x_j \in X$ are adjacent if the length of the geodesic path $\pi_{x_i, x_j}$ in $\mathcal{P} \cup Ext(\mathcal P)$ connecting them is at most $d$ (as we will show soon, computing $G_d(X)$ does not require the shortest-path trees mentioned in Lemma~\ref{lem:max_distance1}). 
We then solve the problem of finding a shortest cycle in $G_d(X)$ from $x_0$ to itself going around $\mathcal P$. 
The set of vertices of that cycle, including $x_0$, is a valid solution to our problem, but not necessarily an optimal one.

Note that the problem of finding a shortest cycle from $x_0$ to itself can be reduced to that of finding a shortest path from $x_0$ to a copy $x'_0=x_r$ of $x_0$. 
To this end, we insert $x'_0$ in $V(G_d(X))$ in such a way that, if the length of the interval $[x_i,x_0]$ is at most $d$, then $x_i$ is adjacent to $x'_0$ instead of $x_0$.

Now we show in detail how the algorithm works, including how to compute~$G_d(X)$.
\vspace{10pt}

\begin{algorithm}[H]
	\DontPrintSemicolon
	\SetAlgoSkip{medskip}
	\KwIn{Polygon $\P$, $s_0\in\P\cap\CH(\P)$, $d>0$, $\epsilon>0$}
	\KwOut{List of stations in a $d$-hull of $\P$}
	
	\nl\If{$s_0 = x_0$ is not a vertex of $\P$}{
	
		\nl{Split the edge containing $s_0$ in two in such a way that $s_0$ becomes a vertex}
		
	}
	
	\nl{Let $V$ be the set of vertices of $\mathcal P$ and set $X=V$}
	
	\nl{Add a copy $x'_0$ of $x_0$ to $X$}
	
	\nl\For{each edge of $\P$ of length $\ell>\epsilon$}{
	
		\nl{Add $\lceil\frac{l}{\epsilon}\rceil$ points to $X$ dividing the edge into segments of length $\le\epsilon$}
		
	}
	
	\nl{Let $X=\{x_0,x_1,\dots,x_{m-1}, x_m=x'_0\}$ be the set of points in clockwise order around $\P$}
	
	\nl{Construct a weighted directed graph $H_d(X)=(X,E)$ with $E=E_1\cup E_2$ defined as follows:
	\begin{enumerate}[label={(\alph*)}]
		\item $(x_i,x_j)\in E_1$ if $j=i+1$ and $x_i,x_{i+1}$ are on the same edge of $\P$
		
		\item $(x_i,x_j)\in E_2$ if $i<j$, and the open segment from $x_i$ to $x_j$ has length $\le d$ and is contained in $\Ext(\P)$
		
		\item The weight of each edge $(x_i,x_j)\in E$ is the Euclidean distance between $x_i$ and $x_j$
	
	\end{enumerate}		
	}
	
	\nl\For{each $x_i\in X$}{
	
		\nl{Use Dijkstra's algorithm to compute $X_i$, the set of vertices of $X$ that can be reached from $x_i$ by a directed path of total weight $\le d$}
		
	}
	
	\nl{Construct a graph with vertex set $X$ where $x_i$ is adjacent to $x_j$ iff $x_j\in X_i$ or $x_i\in X_j$. Since the vertices of $\mathcal{P}$ belong to $X$, one can easily check that this graph is actually $G_d(X)$.}
	
	\nl{Using BFS (or Dijkstra's algorithm with weights 1), compute a shortest path from $x_0$ to $x'_0=x_m$ of minimum length in $G_d(X)$}
	
	\Return {the set of vertices of $G_d(X)$}
	\caption{\AppSol}
	\label{alg:app_sol}
\end{algorithm}
\vspace{20pt}

It is possible to check whether a directed edge $(x_i,x_j)$ belongs to $E_2$ in $O(n)$ time. 
This leads to a total time complexity of $O((\frac{L}{\epsilon})^3 + (\frac{L}{\epsilon})^2 n)$ for \AppSol, where $L$ denotes the total length of $\mathcal{P}$.

This algorithm, while simpler to implement than \OptSol, does not directly yield an approximation to the MinStation problem (this is discussed in more detail in the next section). 
We now show how we can improve on this by applying this algorithm more than once: two applications of the MinStation \AppSol\ algorithm can be used to \emph{certify} the sharpness of a single application of this result, and a logarithmic number of applications can be used to give an additive approximation for MinDistance.

Denote by $\alpha(\mathcal P,s_0,d,\epsilon)$ be the number of base stations found by the \AppSol\ algorithm for given $\mathcal P$, $s_0\in \mathcal P\cap \CH(\mathcal P)$, flight range $d$, and $\epsilon>0$. 
Let $k$ be the minimum number of base stations among all solutions that have $s_0$ as one of their base stations. 
The key is the following result:

\begin{theorem}
\label{thm:approx}
$\alpha(\mathcal P,s_0,d+\epsilon,\epsilon) \leq k\leq\alpha(\mathcal P,s_0,d,\epsilon)$. 
In particular, if $\alpha(\mathcal P,s_0,d+\epsilon,\epsilon) =\alpha(\mathcal P,s_0,d,\epsilon)$, the solution is best possible among those containing $s_0$.
\end{theorem}

\begin{proof}
Clearly, $\alpha(\mathcal P,s_0,d,\epsilon)\ge k$. 
It suffices to show that $\alpha(\mathcal P,s_0,d+\epsilon,\epsilon)\le k$. Consider an optimal set of $k$ base stations $S^*$. 
Let $S$ be a set of base stations obtained by selecting the nearest point in $X$ for each point in $S^*$, then the geodesic distance between consecutive base stations in $S$ is at most $d+\epsilon$.
Therefore $\alpha(\mathcal P,s_0,d+\epsilon,\epsilon)\le k$.
\end{proof}
Since $s_0$ lies on the boundary of the convex hull of $P$, every solution to MinStation must contain a station on a point $z\in\mathcal{P}$ such that $\delta(x_0,z)\leqslant d$. 
This can easily be seen to imply that Theorem 13 can be adapted to work for general solutions (and not only those that contain $x_0$) by modifying the algorithm so that it searches for the shortest path in $G_d(X)$ from $x_i$ to itself for all $x_i$ with $\delta(x_0,x_i)\leqslant d+\frac{\epsilon}{2}$, and then returns the shortest one among all of these. 
This slight variant of \AppSol\ will be called \AppSol 2.

This has immediate implications for MinDistance; if the least number of stations in a solution in $G_{d+\epsilon}(X)$ is at most $k$, then the optimal solution to the MinDistance Problem (find the smallest flight range such that $k$ stations are sufficient) lies between $d$ and $d+\epsilon$.  
Thus by using binary search on $d$, the optimal flight range can be approximated up to an additive constant.

\begin{theorem}
Given a positive integer $k$ and an $\epsilon>0$, it is possible to find a solution to the MinDistance problem using $k$ base stations such that the flight capacity of the drones is at most $\epsilon$ larger than the optimal one. 
This is achieved by running $O(\log|X|)=O(\log(\frac{L}{\epsilon}+n))$ iterations of \AppSol{\em 2} to perform a binary search on the set of all distinct drone (geodesic) distances between pairs of points of $X$.
\end{theorem}

\begin{corollary}
Given a positive integer $k$ and an $\epsilon>0$, 
\begin{itemize}
    \item An additive $\epsilon$-approximation for the MinDistance problem with one fixed base station can be computed in 
$O((n^2 + L n)\log(\frac{L}{\epsilon}+n))$ time.
    \item A quasi-optimal additive $\epsilon$-approximation for the MinDistance problem (i.e. with $k$ or $k+1$ base stations) can be computed in 
$O((n^2 + L n)\log(\frac{L}{\epsilon}+n))$ time.
\end{itemize}
\end{corollary}

\section{Experiments}

We implemented algorithm \AppSol. The program is written in Java and is available at \cite{Reza2022}.
We run experiments on data from Salamis Island using data provided by Harvard WorldMap\footnote{
\href{http://worldmap.harvard.edu}{http://worldmap.harvard.edu}
}.
There are 596 vertices in the polygon representing the island.
The vertices are given by latitude and longitude and we converted them to $(x,y)$ coordinates in meters.
According to our data, the perimeter of the island is 113639.9 meters.

    \begin{table}[htb]
    \begin{center}
    \begin{tabular}{|c|crrr|}
    \hline
    $d$ & $k$ & $\epsilon$ & $T_1$  & $T_2$ \\ 
    \hline
    \hline
    1000  & 85  & 10.48 & 55.502 & 378.434 \\
    1200  & 69  & 12.58 & 40.774 & 360.072 \\
    1250  & 67  & 2.54 & 1034.178 & 7696.363 \\
    1300  & 63 & 1.06 & 6912.250 & 49544.916 \\
    1400  & 54 & 8.49 & 90.388 & 792.151 \\
    1500  & 51 & 15.72 & 36.228  & 317.915  \\
    1700  & 44  & 63.85 & 3.861 & 38.557 \\
    1750  & 43  & 12.74 & 53.491 & 508.694 \\
    1800  & 42  & 2.12 & 2443.456 & 17636.010 \\
    1900  & 38  & 16.60 & 44.015 & 363.716 \\
    2000  & 36 & 52.17 & 7.316 & 73.641 \\
    2100  & 34 & 1.43 & 6772.667 & 49193.534 \\
    2400  & 26 & 90.15 & 2.676  & 39.461  \\
    2500  & 25  & 93.90 & 2.950 & 36.413 \\
    3000  & 20  & 2.45 & 3688.721 & 26533.493 \\
    3200  & 19 & 120.20 & 2.753 & 44.280 \\
    \hline 
    \end{tabular}
    \end{center}
    \caption{
Columns: distance $d$ in meters, $k$ is the optimal number of base stations, 
$\epsilon$ of the last iteration is in meters, 
$T_1$ is the time (in seconds) of the last iteration, and $T_2$ is the total time of \AppSol.}
    \label{bounds}
    \end{table}

\begin{figure}[htb]
    \begin{minipage}{2.5in}
        \begin{center}
            \includegraphics[scale=0.8]{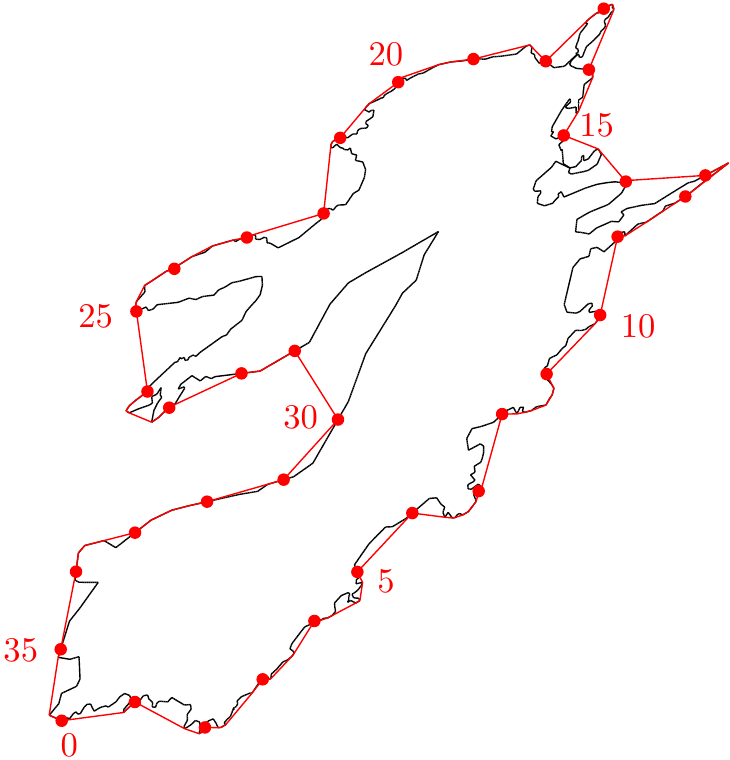}
        \end{center}
    \end{minipage}
    \begin{minipage}{2.5in}
        \begin{center}
            \includegraphics[scale=0.8]{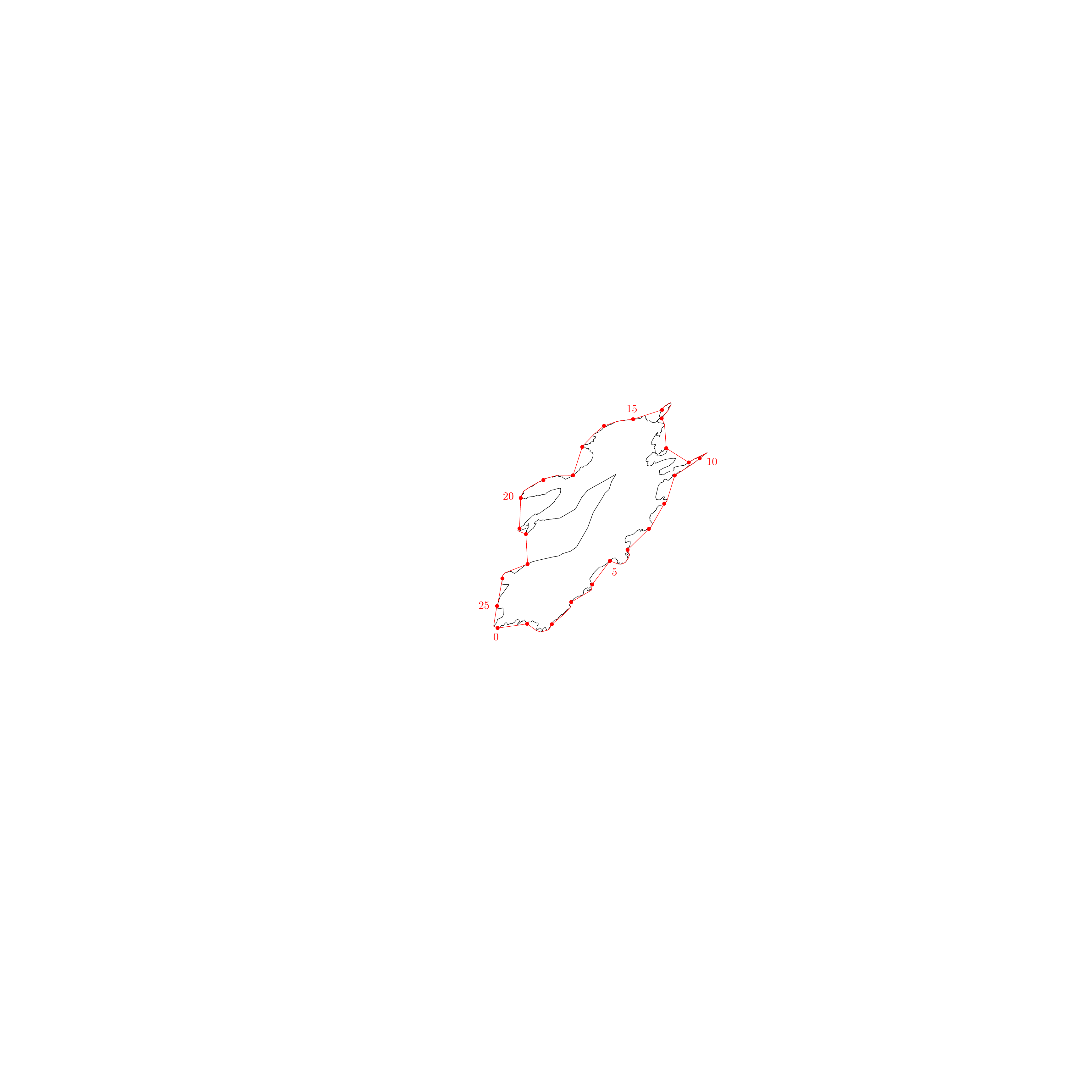}
        \end{center}
    \end{minipage}
    \caption{Salamis Island in the Saronic Gulf.
(a) 36 base stations for $d=2000$. %$d=0.023$.
(b) 26 base stations for $d=2400$.
The number of base stations in (a) and (b) is optimal among those containing $s_0$ by Theorem~\ref{thm:approx}.}
    \label{fig:isl1}
\end{figure}

Our goal was to find the optimal number of base stations using approximation algorithm \AppSol\ and the sufficient condition provided by Theorem~\ref{thm:approx}.
For different values of $d$ and a fixed base station ($s_0 = 0$) on the island, we apply the following approach.
We start with epsilon equal to the drone capacity $d$. 
We divide $\epsilon$ by 1.2 each time it does not satisfy the sufficient condition of Theorem~\ref{thm:approx}. 
Interestingly, the optimal number of base stations was found in all experiments.
The results are shown in Table \ref{bounds}.
The program was executed on a Linux server with 32 core CPUs and 64GB RAM.
Observe that $k$ is monotone with respect to $d$ but $\epsilon$ is not.
Two solutions for $d=2000$ and $d=2400$ are shown in Figure \ref{fig:isl1}.

\section{Conclusions}

In this paper we consider the problem of finding the minimum number of refueling stations along the boundary of an island, modeled as a polygon $\mathcal{P}$ with perimeter $L$, in such a way that a drone with flight range $d$ can follow a polygonal path enclosing $P$. 
We describe an $O(n^2 + \frac{L}{d})$-time algorithm that attains an optimal solution under the restriction that a base station is a point in the intersection of the boundary of $\mathcal{P}$ and its convex hull.
Moreover, if we remove this restriction, our algorithm returns a solution with at most one additional base station with respect to a globally optimal solution.%, which we call a quasi-optimal solution.

The setting of the problem allowed us to suppose that the drones fly at constant height, and therefore the assumption that any drone is always able to fly between base stations at distance at most $d$ is not unreasonable.
However, some applications may require to consider the elevation differences or the presence of obstacles between base stations, which means that re-computing the maximum flight distance each time a base station is placed might be necessary.
If this value can be obtained in linear time per base station, then our algorithm could be adapted for these settings while keeping the original time complexity.
To accomplish this we only have change the value of $d$ in the steps described in Lemma~\ref{lem:projections_2} and Lemma~\ref{lem:max_distance1}, as these steps are done for each base station.

It remains as an open problem to determine if the MinStation problem without the restriction that one base station has to lie on the convex hull can be optimally solved in polynomial time.
This is relevant, since there exist examples in which an optimal solution contains no base station on the convex hull of the island.

We also presented an algorithm to obtain an additive approximation to the problem of minimizing the fuel capacity required for the drones to patrol an island when we are allowed to place at most $k$ base stations around its boundary.
The main tool in this solution is a discretization of the original MinStation problem. %, via an algorithm we call \AppSol. 
This discretized approach also yields an easier to implement algorithm to approximate the MinStation problem, albeit without any theoretical guarantees on the quality of the solution.  
It is also an open problem to determine if an exact solution to the MinDistance problem can be obtained in polynomial time.

\bibliographystyle{plain}
\bibliography{references}

\end{document}